\documentclass[fleqn,usenatbib]{mnras}

\usepackage{txfonts}

\usepackage[T1]{fontenc}

\DeclareRobustCommand{\VAN}[3]{#2}
\let\VANthebibliography\thebibliography
\def\thebibliography{\DeclareRobustCommand{\VAN}[3]{##3}\VANthebibliography}

\usepackage{graphicx}	

\usepackage{amssymb}
\usepackage{hyperref}
\hypersetup{
    colorlinks=true,
    linkcolor=blue,
    filecolor=magenta,      
    urlcolor=cyan,
    pdfpagemode=FullScreen,
    }
\usepackage{booktabs}

\newcommand{\lya}{Ly$\alpha$}
\newcommand{\ha}{H$\alpha$}
\newcommand{\hb}{H$\beta$}
\newcommand{\oiii}{[O\,\textsc{iii}]}
\newcommand{\oii}{[O\,\textsc{ii}]}
\newcommand{\nii}{[N\,\textsc{ii}]}
\newcommand{\neiii}{[Ne\,\textsc{iii}]}
\newcommand{\nev}{[Ne\,\textsc{v}]}
\newcommand{\sii}{[S\,\textsc{ii}]}
\newcommand{\oi}{[O\,\textsc{i}]}
\newcommand{\heii}{He\,\textsc{ii}}

\newcommand{\kms}{km\,s$^{-1}$}

\title[AGN feedback in a radio galaxy at $z=4.1$]{Widespread AGN feedback in a forming brightest cluster galaxy at $z=4.1$ unveiled by JWST}

\author[Saxena, Overzier et al.]{Aayush Saxena,$^{1,2}$\thanks{E-mail: aayush.saxena@physics.ox.ac.uk}$^\dagger$
Roderik A. Overzier,$^{3,4,5}$$^\dagger$
Montserrat Villar-Mart\'{i}n,$^{6}$
Tim Heckman,$^{7}$
\newauthor Namrata Roy,$^{7}$
Kenneth J. Duncan,$^{8}$
Huub R\"{o}ttgering,$^{3}$
George Miley,$^{3}$
Catarina Aydar,$^{9}$
Philip Best,$^{8}$
\newauthor Sarah E. I. Bosman,$^{10,11}$
Alex J. Cameron,$^{1}$
Krisztina \'{E}va Gab\'{a}nyi,$^{12,13,14,15}$
Andrew Humphrey,$^{16,17}$
\newauthor Sandy Morais,$^{17}$
Masafusa Onoue,$^{18,19,20}$
Laura Pentericci,$^{21}$
Victoria Reynaldi,$^{22,23}$
Bram Venemans$^{3}$
\\
$^{\dagger}$These authors contributed equally to this work\\
$^{1}$Department of Physics, University of Oxford, Denys Wilkinson Building, Keble Road, Oxford OX1 3RH, UK\\
$^{2}$Department of Physics and Astronomy, University College London, Gower Street, London WC1E 6BT, UK\\
$^{3}$Leiden Observatory, University of Leiden, Niels Bohrweg 2, 2333 CA Leiden, The Netherlands\\
$^{4}$Observat\'orio Nacional/MCTI, Rua General Jos\'e Cristino 77, Rio de Janeiro, RJ 20921-400, Brazil\\
$^{5}$TNO, Oude Waalsdorperweg 63, 2597 AK, Den Haag, The Netherlands\\
$^{6}$Centro de Astrobiolog\'{i}a (CAB), CSIC-INTA, Ctra. de Ajalvir, km 4, 28850 Torrej\'{o}n de Ardoz, Madrid, Spain\\
$^{7}$Center for Astrophysical Sciences, Department of Physics and Astronomy, Johns Hopkins University, Baltimore, MD, 21218, USA\\
$^{8}$Institute for Astronomy, University of Edinburgh Royal Observatory, Blackford Hill, Edinburgh, EH9 3HJ, UK\\
$^{9}$Max-Planck-Institut f\"ur Extraterrestrische Physik, Gie{\ss}enbachstra{\ss}e, D-85748 Garching, Germany\\
$^{10}$Institute for Theoretical Physics, Heidelberg University, Philosophenweg 12, D-69120, Heidelberg, Germany\\
$^{11}$Max-Planck-Institut f\"ur Astronomie, K\"onigstuhl 17, D-69117, Heidelberg, Germany\\
$^{12}$Department of Astronomy, Institute of Physics and Astronomy, ELTE Eo\"tov\"os Lor\'and University, P\'azm\'any P\'eter S\'et\'any 1/A, H-1117, Budapest, Hungary\\
$^{13}$HUN-REN–ELTE Extragalactic Astrophysics Research Group, ELTE Eo\"tov\"os Lor\'and University, P\'azm\'any P\'eter S\'et\'any 1/A, H-1117, Budapest, Hungary\\
$^{14}$Konkoly Observatory, HUN-REN Research Centre for Astronomy and Earth Sciences, Konkoly Thege Mikl\'os \'ut 15-17, H-1121 Budapest, Hungary\\
$^{15}$CSFK, MTA Centre of Excellence, Konkoly Thege Mikl\'os \'ut 15-17, H-1121 Budapest, Hungary\\
$^{16}$DTx -- Digital Transformation CoLab, Building 1, Azur\'em Campus, University of Minho, 4800-058 Guimar\~aes, Portugal\\
$^{17}$Faculdade de Ci\^{e}ncias da Universidade do Porto, Rua do Campo de Alegre, 4150-007, Porto, Portugal\\
$^{18}$Kavli IPMU, WPI, The University of Tokyo, 5-1-5 Kashiwanoha,
Kashiwa, Chiba 277-8583, Japan\\
$^{19}$Center for Data-Driven Discovery, Kavli IPMU (WPI), UTIAS, The University of Tokyo, Kashiwa, Chiba 277-8583, Japan\\
$^{20}$Kavli Institute for Astronomy and Astrophysics, Peking University, Beijing 100871, P.R.China\\
$^{21}$INAF – Osservatorio Astronomico di Roma, via Frascati 33, 00078, Monteporzio Catone, Italy\\
$^{22}$Instituto de Astrof\'isica de La Plata, CONICET-UNLP, Paseo del Bosque, B1900FWA La Plata, Argentina\\
$^{23}$Facultad de Ciencias Astron\'omicas y Geofísicas, Universidad Nacional de La Plata, Argentina
}

\date{Accepted XXX. Received YYY; in original form ZZZ}

\pubyear{2024}

\begin{document}
\label{firstpage}
\pagerange{\pageref{firstpage}--\pageref{lastpage}}
\maketitle

\begin{abstract}
We present rest-frame optical spectroscopy using \emph{JWST}/NIRSpec IFU for the radio galaxy TN J1338$-$1942 at $z=4.1$, one of the most luminous galaxies in the early universe with powerful extended radio jets. Previous observations showed evidence for strong, large-scale outflows on the basis of its large ($\sim$150 kpc) halo detected in \lya, and high velocity \oii\ emission features detected in ground-based IFU data. Our NIRSpec/IFU observations spatially resolve the emission line properties across the host galaxy in great detail. We find at least five concentrations of line emission, coinciding with discrete continuum features previously detected in imaging from \emph{HST} and \emph{JWST}, over an extent of $\sim2$\arcsec\ ($\sim15$ kpc). The spectral diagnostics enabled by NIRSpec unambiguously trace the activity of the obscured AGN plus interaction between the interstellar medium and the radio jet as the dominant mechanisms for the ionization state and kinematics of the gas in the system. A secondary region of very high ionization lies at roughly 5 kpc distance from the nucleus, and within the context of an expanding cocoon enveloping the radio lobe, this may be explained by strong shock-ionization of the entrained gas. However, it could also signal the presence of a second obscured AGN, which may also offer an explanation for an intriguing outflow feature seen perpendicular to the radio axis. The presence of a dual SMBH system in this galaxy would support that large galaxies in the early Universe quickly accumulated their mass through the merging of smaller units (each with their own SMBH), at the centers of large overdensities. The inferred black hole mass to stellar mass ratio of $0.01 - 0.1$ for TNJ1338 points to a more rapid assembly of black holes compared to the stellar mass of galaxies at high redshifts, consistent with other recent observations. 
\end{abstract}




\section{Introduction}
\label{sec:introduction}
Observations of luminous active galactic nuclei (AGN) in the early Universe show that they are hosted by massive galaxies harbouring active supermassive black holes (SMBHs). The space density of these luminous AGN peaks at $z\sim2$ \citep{rigby15,williams18}, long before the rapid rise of fainter AGN between $z\sim2$ and $z\sim0$ \citep{enoki14}. This downsizing trend in AGN evolution is due to a biased growth scenario, in which the more massive galaxies and their SMBHs were already in place at $z>2$ \citep{merloni04}. The large black hole masses required to power the radio luminosities of radio galaxies and quasars suggest that these early SMBHs grew rapidly through near- (or super-) Eddington accretion \citep{mclure04,netzer07}. It is widely believed that such rapid SMBH accretion can play a vital role during the growth of the host galaxy through the process of AGN ``feedback'' \citep[e.g.][]{fabian12,kormendy13,heckman14,fan23}. High redshift radio galaxies \citep[HzRGs;][]{mccarthy93,miley08} are (obscured) hosts of luminous extended radio sources at $z\gtrsim2$, powered by presumably efficient thin-disk accretion around rapidly spinning, massive SMBHs \citep{blandford77,rees84}. The large sizes of the jets imply long-term AGN activity with strong kinetic forms of feedback besides radiative feedback \citep[e.g.][]{saxena17}. Since HzRGs are often embedded in dense environments such as massive groups or protoclusters, they are believed to be progenitors of the massive central galaxies in clusters today \citep[e.g.][]{hatch09,chiaberge11,emonts16,overzier16,retana-montenegro17,nesvadba17a,sabater19,hardcastle20,magliocchetti22,poitevineau23}. 

The nuclear obscuration in HzRGs provides a natural screen to ensure that the nuclear emission does not outshine the host galaxy at UV/optical wavelengths \citep{miley08}. This allows direct access to the starlight, interstellar medium (ISM) and (spatially resolved) narrow line regions and thus the interaction of AGN-related phenomena with stars, gas and dust. Some fraction of HzRGs that are completely obscured in the rest-UV do, however, show broad Balmer lines in the rest-optical \citep{nesvadba11}, thus also giving a way to infer the black hole masses, similar to unobscured quasars. Combined with their extended continuum and nebular emission that are interacting with the radio jets, HzRGs are thus unique laboratories for studying the active growth phase of massive galaxies and their SMBHs that are the progenitors of massive galaxies on the local relation between black hole mass and stellar velocity dispersion \citep[the $M_{\rm{BH}}-\sigma_*$ relation;][]{kormendy13}. 

Extended emission line nebulae with sizes approaching $\sim$0.5 Mpc that are bright in Ly$\alpha$ and other lines were first discovered around radio galaxies \citep{mccarthy87,heckman91}. Especially images from the \emph{HST}, supplemented by ground-based spectroscopy probing the properties of the nebular gas, have shown that HzRGs possess some of the most remarkable host galaxy morphologies among {\it all} galaxies known at $z\gtrsim2$, with e.g. (1) complex merger systems, (2) star formation visibly aligned with radio jets, (3) radio jets bending in or near the host \citep[e.g.][]{chambers90, dey97, miley06, miley08, drouart16, pentericci98, pentericci99, pentericci01, miley06, saxena18b, saxena19, duncan23}. Also, HzRGs have exceptionally large bolometric luminosities, and they are found at the extreme end of the $L_{\scriptsize{\textrm{[O~\textsc{iii}]}}}-L_{\scriptsize{\textrm{radio}}}$ relation where few sources in the local Universe are found \citep{xu99}. 

Resolved studies of the ionisation, metal abundance and kinematics using slit or integral field unit (IFU) observations at optical and near-IR wavelengths have highlighted the important role of both jet-induced turbulence and radiation-driven feedback that dominate the energy budget of the ISM and thus the future evolution of these massive galaxies \citep{best00,villar-martin07a,nesvadba11,nesvadba17b}. These studies rely on the important rest-frame optical emission lines, enabling line ratio diagnostics  \citep[BPT;][]{baldwin81} that can differentiate between the sources of the ionising radiation powering the lines (AGN versus star-formation), determine the ISM metallicity, and density (e.g., using the [S \textsc{ii}] diagnostic). Alignment of the line-emitting gas and variability of line strengths and ratios along radio hotspots \citep{vanojik96} have also pin-pointed the crucial role of radio jets in driving out gas and quenching star-formation \citep[e.g.][]{nesvadba06,morais17,falkendal19,kolwa19,man19,wang2023}.

Previous work on HzRGs mainly based on \emph{HST} photometry and seeing-limited ground IFU observations did not allow to probe the ISM on similar scales as the radio jets, and cleanly separate the nuclear, stellar and ISM properties of the host galaxies \citep[see also, however][]{wang2023}. As a result, important open questions still remain surrounding how exactly AGN feedback is triggered, how radio jets interact with their dense surroundings and how this affects the growth of massive galaxies and their SMBHs. \emph{JWST} has made it possible, for the first time, to constrain the role of jet-gas interactions and AGN feedback in regulating the mass, size, kinematics and physical conditions of forming massive galaxies.

In this paper we present \emph{JWST}/NIRSpec IFU observations of the radio galaxy TN J1338--1942 at $z = 4.11$ \citep{debreuck99,pentericci00}. Ground-based imaging and spectroscopy have revealed a large-scale \lya\ halo extending to $\sim$150 kpc \citep{debreuck99,debreuck00a,debreuck00b,venemans02,villar-martin07b}. Spatially resolved \lya\ spectra showed evidence for high-velocity, large-scale outflowing gas of large mass \citep{swinbank15}, while imaging with \emph{HST} and \emph{JWST} showed a complex and very extended host galaxy consisting of several bright clumps \citep{miley04,zirm05,overzier08,duncan23}. The overall \lya\ and UV/optical morphologies are aligned with the radio axis. \citet{zirm05} found evidence of a wedge-shaped \lya\ feature perpendicular to the radio axis that could be outflow or scattered light. 

Ground-based IFU observations detected \neiii\ and \oii\ emission across the host galaxy with maximum velocity offsets and velocity dispersions of, respectively, 500 and 1500 \kms\ \citep{nesvadba17a}. Photometry from \emph{HST}, \emph{Spitzer} and \emph{JWST} have shown that with a stellar mass of $\sim10^{11}$ $M_\odot$ TN J1338--1942 is among the most massive galaxies at $z\sim4$ \citep{overzier09,duncan23}. However, previous findings of very high SFRs in relation to radio jet triggering that were based on photometry from \emph{HST}/ACS and \emph{JWST}/NIRCam \citep{zirm05,duncan23} will be invalidated by the results presented in this paper. 

The radio galaxy lies at the center of a dense region of the cosmic web or protocluster traced by overdensities of star-forming galaxies \citep{venemans02,venemans07,miley04,debreuck04,overzier08,saito15}. Combined with the ample evidence for SMBH activity, this makes TN J1338--1942 one of the best known examples of a brightest cluster galaxy progenitor. TN J1338--1942 was targeted because of the rich available multi-wavelength data coupled with unique morphological and kinematic signatures of the gas in and around the galaxy that make it an ideal candidate to directly investigate the impact of AGN feedback with the NIRSpec IFU. In this paper we present a first look into these NIRSpec/IFU observations, while another paper \citep{roy24} investigates in detail the outflows and gas kinematics in this system.

The layout of this paper is as follows. In Section \ref{sec:observations} we present the observations and data reduction methods employed. In Section \ref{sec:results} we present the main results of this finding, which include 1D spectra and 2D line maps, gas kinematics and ionization state in this system. In Section \ref{sec:discussion} we present a discussion on what these new observations mean for the nature of TNJ1338$-$1942, assessing scenarios that may explain its observed properties. Finally, in Section \ref{sec:summary} we summarize the findings of this study.

Throughout this paper, we assume flat $\Lambda$CDM cosmology following \citet{planck20} (included in \textsc{astropy.cosmology}; $H_0 = 67.7$ km s$^{-1}$ Mpc$^{-1}$, $\Omega_m = 0.31$). This gives a spatial scale of $7.0$ proper kpc\,arcsec$^{-1}$ at $z=4.1$. 

\section{Observations and Data Reduction}
\label{sec:observations}

The \emph{JWST} spectroscopic observations of TN J1338$-$1942 (RA: 13:38:26.1, Dec: -19:42:31.4, henceforward ``TNJ1338") presented in this paper (ID: GO1964, PIs: Overzier and Saxena) were taken on 22 February 2023. Two high-resolution ($R\sim2700$) grating/disperser combinations were used, F170LP/G235H and F290LP/G395H. For both grating/disperser combinations a 4-POINT-DITHER pattern was used. A total of 18674 seconds of exposure time was obtained in each of the two grating/disperser combinations. Additionally, two background exposures using exactly the same setup and exposure times were taken to allow for background subtraction of the science data cube using a completely source-free region near the science target.

For the most part, pipeline steps from the official data reduction pipeline released by Space Telescope Science Institute\footnote{\url{https://jwst-pipeline.readthedocs.io/en/stable/jwst/introduction.html}} were used to reduce and calibrate the raw data. We implemented a number of additional steps during the reduction process to improve the quality of the final reduced data products, which we highlight below.

We first downloaded all Stage 1 calibration products (the rate files) from the Mikulski Archive for Space Telescope (MAST) Portal\footnote{\url{https://mast.stsci.edu/portal/}}, which were already gain-corrected, flagged for snowballs and jumps, as well as bias and dark subtracted. Following \cite{mar23}, the zero-levels of the science dither frames after Stage 1 processing were made consistent by subtracting the median value from each 2D image. Stage 2 of the data reduction pipeline was then performed on these zero-level corrected Stage 1 products, which includes world coordinate system (WCS) assignment, background and imprint subtraction, flat-field correction, pathloss correction, photometric calibration and distortion correction.

Crucially, we opted not to use the dedicated background exposures for background subtraction, mainly to avoid adding extra sources of noise into the final science cube. \citet{welch23} showed that the impact on the continuum subtracted line fluxes is minimal when not choosing a dedicated background exposure, and since the main aim of this paper is to explore the strong emission lines from this system, we avoid using the background exposure. Therefore, the background was estimated using the sky spaxels from within the science cube.

At the end of Stage 2, following \cite{mar23} we implemented a custom outlier/cosmic ray detection and flagging routine based on the \textsc{python} version of the \textsc{lacosmic} algorithm\footnote{\url{https://lacosmic.readthedocs.io/en/stable/}} \citep{lacosmic} on all individual 2D calibrated (or cal) images. After flagging outliers, we proceeded with the Stage 3 processing that combined all of the individual exposures into a final, fully calibrated 3D data cube, using the ``Shephard's method'' in the pipeline (\textsc{weighting=emsm}). The final data cubes were then visually inspected and any remaining outliers or cosmic ray residuals were identified and flagged.

\subsection{Astrometric accuracy}

The small $3''\times3''$ field of view of the NIRSpec IFU means that absolute astrometric calibrations cannot be explicitly performed. Therefore, we verified our astrometry by comparing the distribution of the line and continuum flux in our cube with existing NIRCam images from \citet{duncan23}, where the NIRCam astrometry was registered using Gaia DR3 and existing Subaru data in the field, leading to a $\sim2$\,mas accuracy. We convolved our cube with NIRCam filters to generate synthetic NIRCam images that could then be visually lined up with the NIRCam data and allowing a qualitative estimate of the astrometric accuracy of the IFU data similar to that of the NIRCam images, and found that the astrometry in our IFU cube and the NIRCam images agreed within the each others' uncertainties.

\subsection{Masking}

Regions of relatively high noise arise at the edges of the data cube because of the adopted 4-POINT dither pattern. Additionally, several cosmic ray residuals remain in the final data cube, particularly towards the edges. To aid the analysis and visualization of continuum and emission line flux, we employed a highly conservative masking procedure. The mask is created using the \oiii\ emission line map, with all pixels lying below $2\sigma$ in \oiii\ significance masked out. We ensured that no continuum emission was inadvertently masked out, by creating continuum and emission line maps from the full data cube and comparing the distributions. Indeed, the spatial extent of emission is largest in the \oiii\ line, which means that masking out lower \oiii\ emission significance spaxels does not mask any continuum flux from this system. Future work will explore possible fainter emission features around this object.

\section{TNJ1338: a complex radio galaxy at $\mathbf{z=4.1}$}
\label{sec:results}
\begin{figure*}
    \centering
    \includegraphics[width=\linewidth]{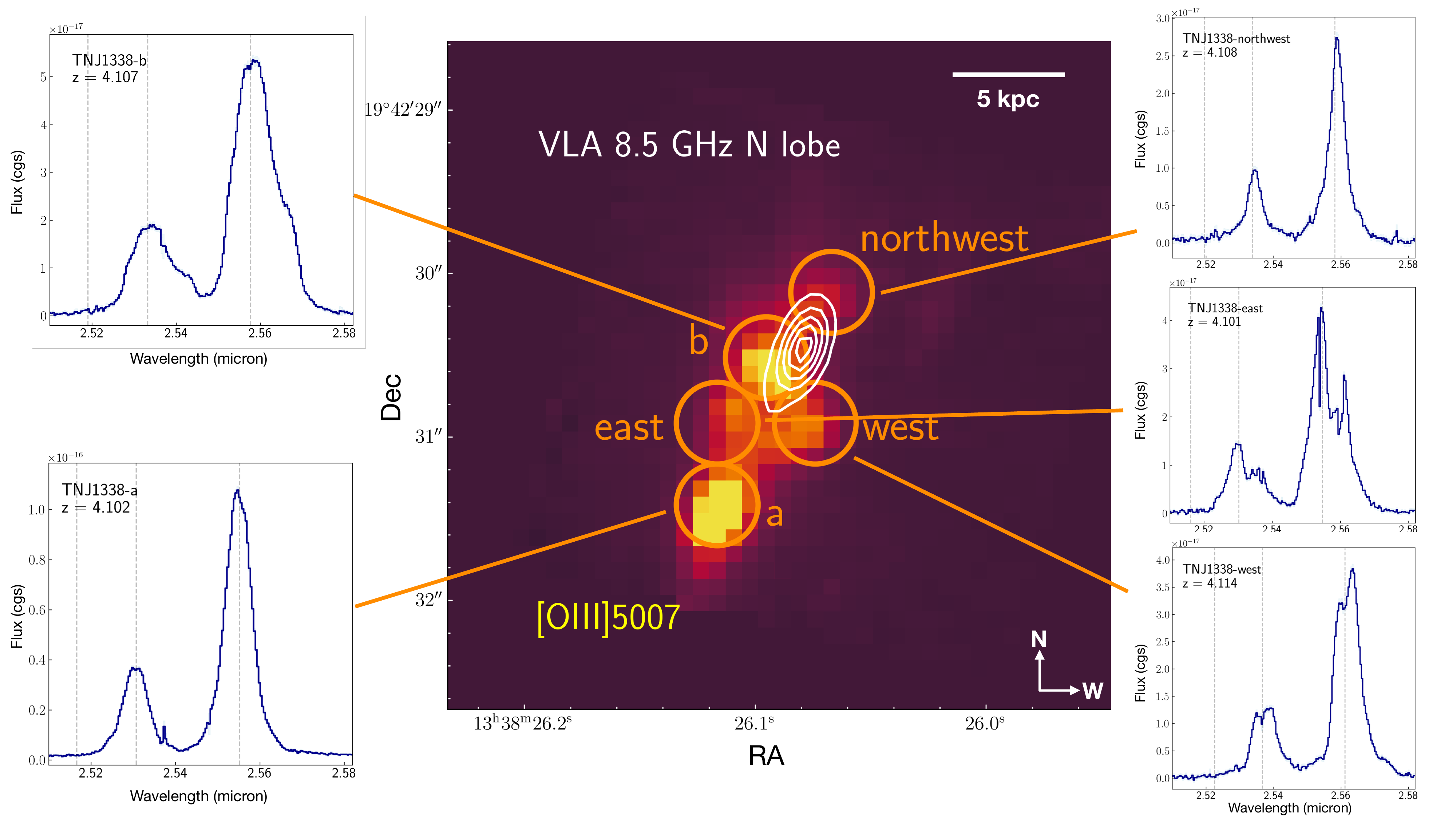}
    \caption{\oiii\,$\lambda5007$ 2D image of TNJ1338, where the VLA 8.5GHz radio contours showing the northern radio lobe \citep{pentericci00} have been overlaid (white) along with 2.5 spaxel radius apertures placed on individual strong line emitting regions (orange). Zoom-ins of the \oiii\,$\lambda\lambda4959, 5007$ lines at the locations of these regions are also shown, highlighting that each line emitting region is at a slightly different redshift, as well as the complex kinematic structure of the ionized gas in this system.}
    \label{fig:TNJ-overview}
\end{figure*}

In this section we discuss the observed spectroscopic properties of TNJ1338. Although previous observations already highlighted the complex host galaxy morphology of TNJ1338 \citep{miley04,zirm05,duncan23}, the NIRSpec/IFU observations reveal the extent of this complexity both in the spatial dimensions and in the kinematics and ionization state of the gas in the system. Figure \ref{fig:TNJ-overview} shows the full (unmasked) two-dimensional \oiii\,$\lambda\lambda 4959,5007$ emission line map for the system, along with zoom-ins into the strong line emitting regions that we have identified in the data cube, which we present in the section that follows. 

The \oiii\ emission extends over about 2\arcsec\ and consists of a number of bright clumps amidst more diffuse emission. Radio contours from the VLA 8.46\,GHz emission from \citet{pentericci00} are shown overlaid. This radio feature is believed to represent the location of the Northern radio lobe and hotspot synchrotron emission. Although a radio core has not been detected, the SMBH powering the radio jet is believed to be associated with the component labeled `a' (see discussion below). This would imply that an invisible radio jet lies roughly along the line connecting component `a' and the bright region detected with the VLA.

\subsection{Identification of line emitting regions and 1D spectra}
\label{sec:regions}
\begin{figure*}
    \centering
    \includegraphics[width=\linewidth]{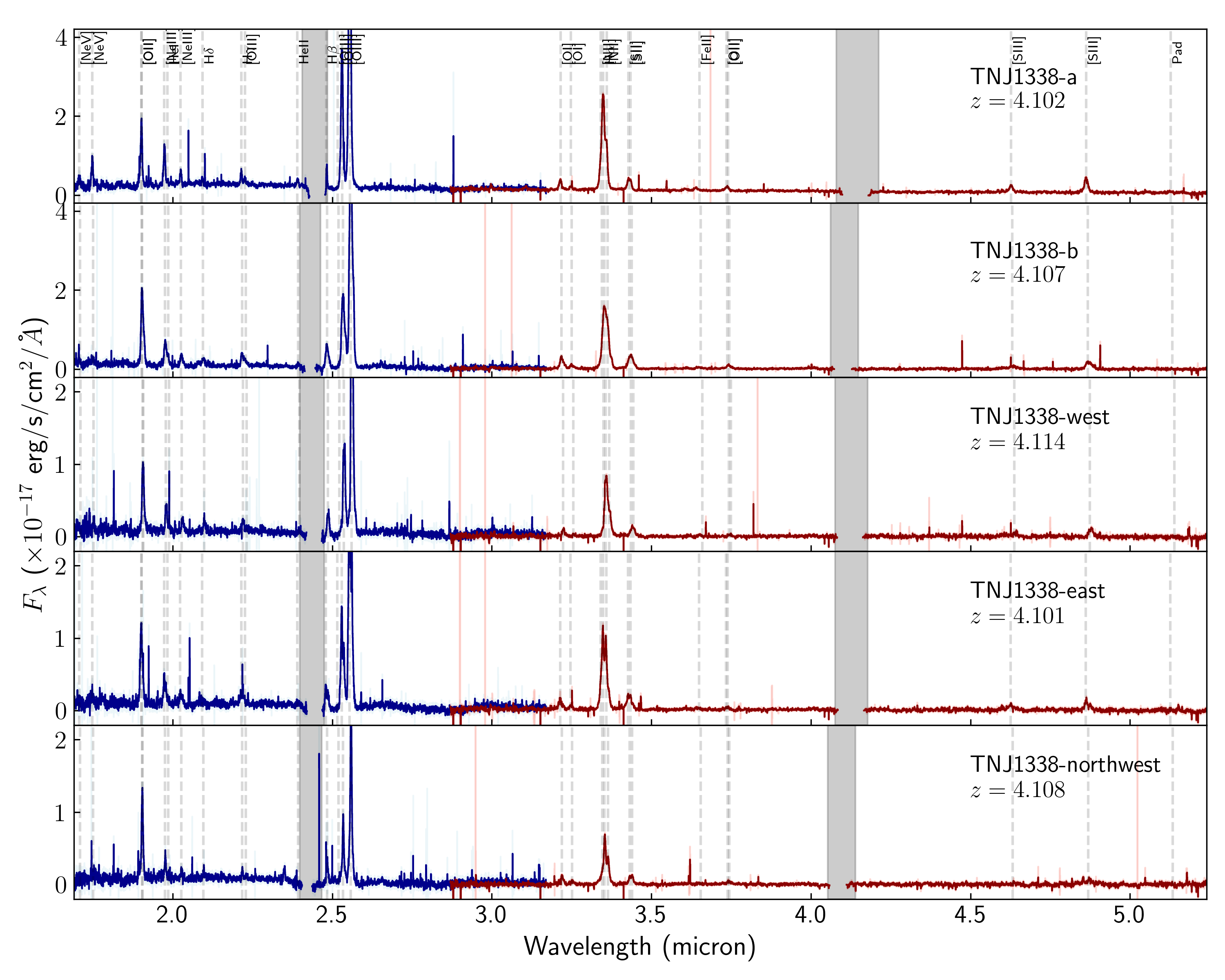}
    \caption{Extracted full 1D spectra from G235H and G395H gratings of all of the strong line-emitting regions identified in the system. A range of high and low ionization emission lines are detected across all of these regions that have been marked with dashed lines, with the most prominent line being \oiii. The shaded grey regions represent detector chip gaps. These spatially resolved line detections enable new insights into the nature of TNJ1338.}
    \label{fig:TNJ-1dspec}
\end{figure*}

In total, we identify five distinct line emitting regions in the system. We place circular apertures with a radius of 2.5 spaxels, which are shown in Figure \ref{fig:TNJ-overview}, with a zoom-in on the region around the \oiii\ emission line. The full extracted 1D spectra from the F170LP/G235H (blue) and F290LP/G395H (red) disperser/grating combinations covering the wavelength range $\approx$1.8-5.2 $\mu$m continuously are shown in Figure \ref{fig:TNJ-1dspec}, with the two narrow intervals at $\approx$2.5 and 4.1 $\mu$m arising due to the detector chip gap indicated by the grey-shaded regions. The levels of the continuum across both setups agree well, suggesting that the data reductions and flux calibrations are internally consistent.  

The component that coincides with the galaxy nucleus, and the presumed site of the SMBH powering the radio emission, is labeled `a' (bottom panel). The second brightest region in \oiii\ is labeled `b' (second panel from the top). We further identify a region `west' (third panel from the top, roughly coinciding with the apex of the previously identified \lya\ `wedge' from \citealt{zirm05}), and `east' (fourth panel from the top), which lies on the opposite side of the radio jet axis. Finally, we identify a region at the north-western edge of the source (`northwest'). It is interesting to note that the centroid of the radio hotspot does not appear to coincide with any region in particular, but appears to lie between `b' and `northwest', with a slight offset in the direction of `west'. 

We measure redshifts from the extracted 1D spectra for all regions by calculating the median of the observed peak of multiple strong emission lines (mainly \ha, \oiii, \hb, \oii). Interestingly, note that the redshifts of each of the line-emitting components are different. TNJ1338-a lies at $z=4.102$, while the second brightest line-emitting region, TNJ1338-b lies at $z=4.107$ with another line emitting component at an even higher redshift, as discussed later in the paper. TNJ1338-northwest that is the furthest away from the nucleus has a redshift $z=4.108$. The west component corresponding to the \lya\ wedge has the highest redshift with $z=4.114$. The east component has the lowest redshift with $z=4.101$. Taking the redshift of TNJ1338-a as the systemic redshift of the host galaxy, this implies relative velocities between the components of $+411$\,\kms\ (TNJ1338-b), $+294$\,\kms\ (northwest), $+647$ \,\kms\ (west), and $-118$\,\kms\ (east). 

Although it is difficult to interpret these velocity differences in the presence of possible outflows, rotation and merging structures, the velocity difference of $\sim750$\,\kms\ between the smallest and largest value is much larger than a gravitationally induced circular velocity expected even for a system as massive as TNJ1338. A closer look at the kinematic properties as traced by the \oiii\ emission is presented in Section \ref{sec:kinematics}, and in more detail including modeling in \citet{roy24}.

The above results show that the already complex morphology of TNJ1338 is further complicated by its highly non-uniform emission line kinematics. The next sections will take a closer look at the kinematic properties and ionization states across the system.  

\subsection{2D emission line intensity maps}

\begin{figure*}
    \centering
    \includegraphics[width=\textwidth]{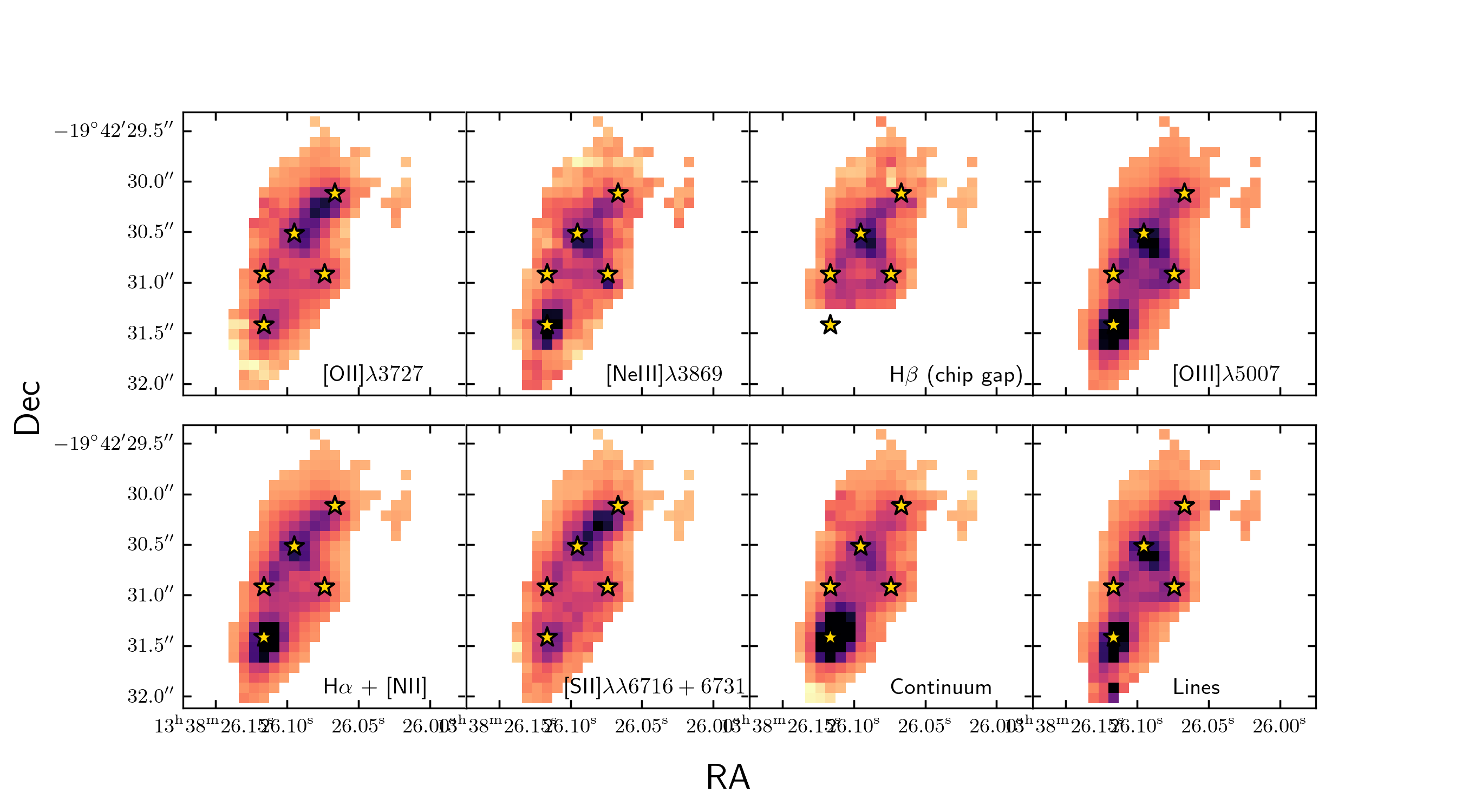}
    \caption{2D line maps, in increasing order of wavelength, showing \oii\,$\lambda3727$, \neiii$\lambda3869$, \hb\ (the coverage in the lower part of the cube, including at TNJ1338-a, falls in the chip gap), \oiii\,$\lambda5007$, \ha+\nii, \sii$\lambda\lambda6716,6731$, continuum in G235H and line emission in G235H, with a linear scale and dynamic range from $-2\sigma$ to $5\sigma$ for each map. The locations of the central spaxels of all five individual line emitting regions are marked using gold stars. The distribution of high and low ionization line emission in this system is important to establish the dominant sources of ionization in TNJ1338, as discussed in this paper.}
    \label{fig:linemaps}
\end{figure*}

In Figure \ref{fig:linemaps} we show the 2D line intensity maps for a variety of common and high-ionization emission lines that will be used in this paper to shed light on the complex ionized gas distribution in the system. These lines include \ha+\nii, \hb, \oi\,$\lambda6300$, \oii\,$\lambda3727$, \oiii\,$\lambda5007$, and \neiii\,$\lambda3869$. The line maps have been created by calculating the mean flux in the data cube within a 100 \AA\ wavelength range around the expected wavelength of each line (with an assumed line peak at $z=4.104$) and subtracting the continuum. Unfortunately, in some locations the \hb\ line falls in the IFU detector gap, rendering the lower part of the datacube at the \hb\ wavelength unusable. 

The last two panels show maps of the summed line-free continuum and of the stacked line emission in the G235 spectrum. Some immediate observations are the relatively high intensities of components TNJ1338-a and TNJ1338-b in the higher ionization lines, a region of relatively strong lower ionization lines (\oi, \oii, \sii) that dominates in the North beyond TNJ1338-b, and the relatively high intensity of the continuum centered on component TNJ1338-a.    

\subsection{Kinematics in TNJ1338a and TNJ1338b}
\label{sec:kinematics}

In order to assess the complexity of the emission line profiles that our line-fitting code will need to capture, we investigate the two bright line emitting regions TNJ1338-a and TNJ1338-b, in particular the \ha\ + \nii\ complex and the \oiii\,$\lambda \lambda 4959,5007$ doublet. Due to the presence of the AGN and the possibility of outflows and jet-ISM interactions, the best-fitting line profile will likely require contributions from several narrow or broad components of independent central velocities. For the line fitting, the local continuum was measured by taking the median continuum value blueward and redward of the emission line and then subtracted. The line fits are performed using Gaussian functions via the non-linear optimization and curve fitting routine \textsc{lmfit}\footnote{\url{https://lmfit.github.io/lmfit-py/}}. We also note here that the observed line shapes are much broader than the instrumental line spread function (LSF $\sim100$\,\kms), and therefore, the LSF is not expected to substantially impact the measured line widths.

\begin{figure*}
    \centering
    \includegraphics[width=0.48\textwidth]{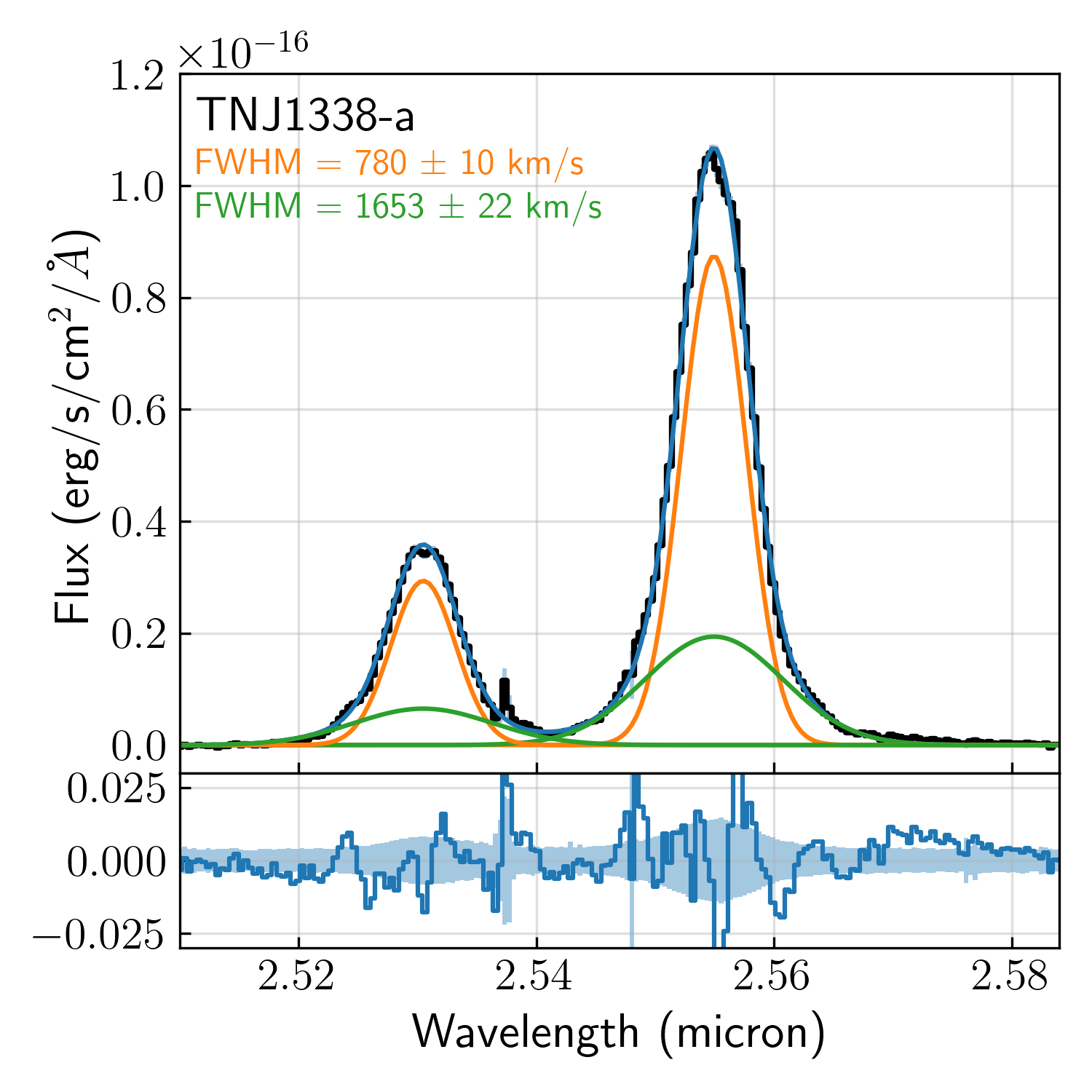}
    \includegraphics[width=0.48\textwidth]{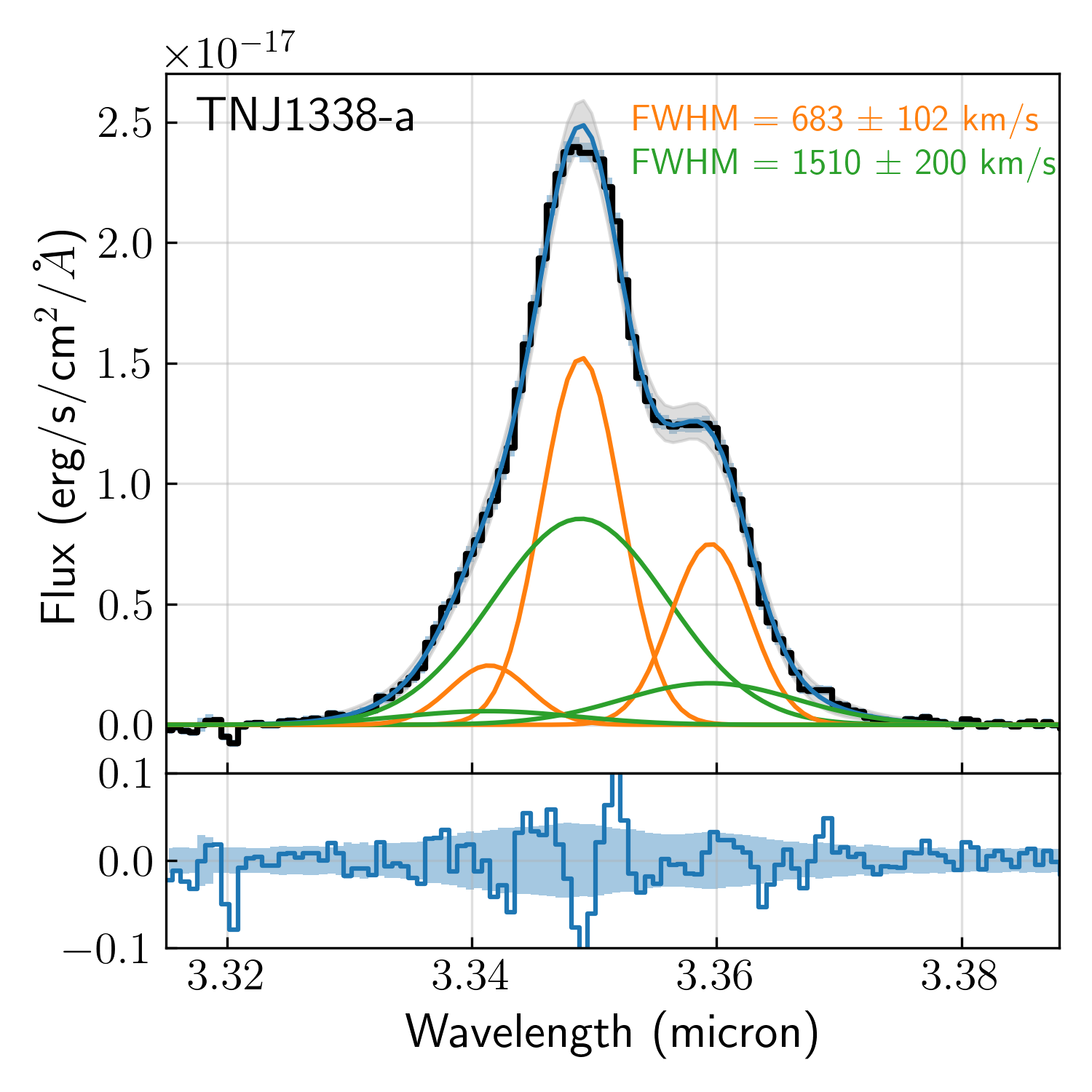}
    \caption{Fits to the \oiii\ (left) and \ha\ + \nii\ (right) lines for TNJ1338-a. The grey shaded region on the top panel represents the best-fitting model uncertainty, whereas the blue shaded region in the bottom panel represents the uncertainty in the data compared to the residuals. We note the presence of narrow and broad components with comparable widths in both these lines, which suggests that the broadening of the lines is driven by outflows and not the presence of an obvious broad-line region (BLR).}
    \label{fig:tnj-a_linefits}
\end{figure*}

\smallskip
\noindent
{\it TNJ1338-a:} The fits to the \oiii\ and \ha+\nii\ lines in TNJ1338-a are shown in Figure \ref{fig:tnj-a_linefits}. For the \ha+\nii\ complex, the fitting is performed using two Gaussian functions, one tracing the broad component and the other narrow, for both the permitted \ha\ line and the forbidden \nii\ doublet. We use a fixed \nii\,$\lambda6583/\lambda6548$ doublet flux ratio of $3.049$. A similar two component Gaussian fitting is implemented for the \oiii\ doublet with the 5007\AA\ to 4959\AA\ flux ratio fixed to 2.98. The results of the fitting did not improve when allowing for separate redshifts of the narrow and the broad components, and these were set to be the same for each line fit. 

As can be seen from Figure \ref{fig:tnj-a_linefits}, the \ha+\nii\ complex requires broad components for both \ha\ and \nii, with the broad component (green) having FWHM $=1510\pm200$\,\kms\ and the narrow component (orange) having FWHM $=683\pm102$\,\kms\ (FWHM). These values are consistent with those obtained for the \oiii\ doublet, with a broad line FWHM $=1653 \pm 22$\,\kms\ and a narrow line FWHM $=780 \pm 10$\,\kms\ (FWHM). From these results, two interesting conclusions can be derived about TNJ1338-a. First, the fact that the full line profiles are so similar between the permitted \ha\ line and the forbidden \nii\ and \oiii\ doublets rules out that the broad line component traces an obvious broad line region (BLR) associated with a SMBH in TNJ1338-a. Second, the fact that the broad and narrow components lie at the same centroid velocity indicates the clear presence of outflowing gas from the location of TNJ1338-a.

\begin{figure*}
    \centering
    \includegraphics[width=0.48\textwidth]{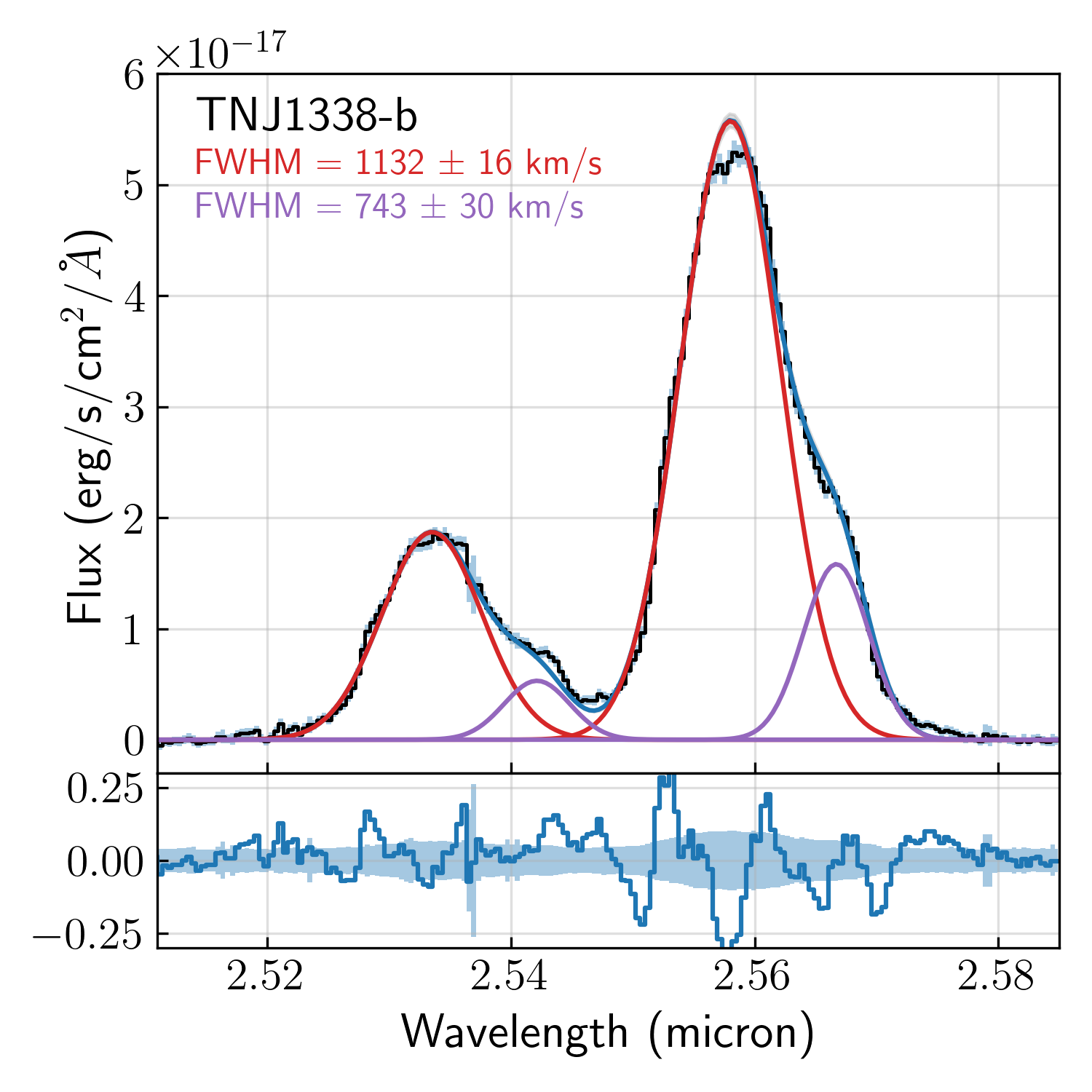}
    \includegraphics[width=0.48\textwidth]{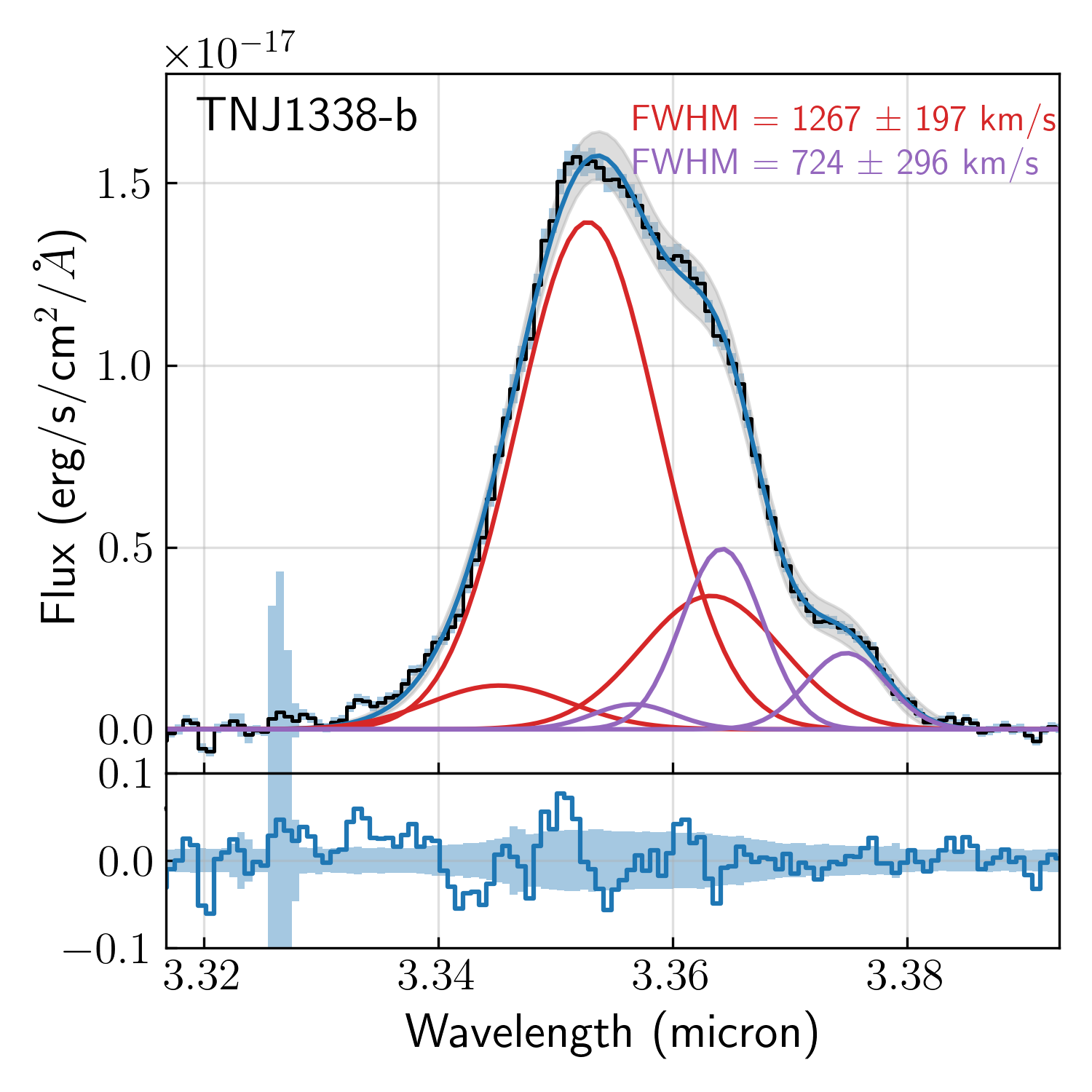}
    \caption{Fits to the \oiii\ (left) \ha\ + \nii\ (right) lines for TNJ1338-b. Here, we fit two independent components to the lines, allowing the redshift of each component to vary. This was needed to accurately fit the highly complex line profiles seen. The redshifts and FWHM of each of the two components resulting in the best-fitting profiles are the same for the \ha+\nii\ complex and the \oiii\ doublet, describing the emission arising at the location of TNJ1338-b well.}
    \label{fig:tnj-b_linefits}
\end{figure*}

\smallskip
\noindent
{\it TNJ1338-b:} We repeat the same line-fitting analysis for the extracted 1D spectrum of TNJ1338-b. As can be immediately seen from Figure \ref{fig:tnj-b_linefits}, fitting two components to each line at the same redshift would not yield an acceptable fit. Therefore, the lines are fitted using two independent velocity components at different redshifts. The best-fitting line profiles for the \ha+\nii\ complex and the \oiii\ doublet are shown in Figure \ref{fig:tnj-b_linefits}.  

We find that the broader line component in \oiii\ has a FWHM $=1132\pm16$\,\kms, and the narrower component has FWHM $=743\pm30$\,\kms. The widths of the components in the \ha+\nii\ complex are once again highly comparable to that of \oiii, with broad FWHM $=1267\pm197$\,\kms\ and narrow FWHM =$724\pm296$\,\kms. This once again rules out the presence of any BLR at this location, and is indicative of outflows, albeit perhaps not originating at the same location as the narrow line emitting region.

Interestingly, at the location of TNJ1338-b the broader line component is brighter and peaks at a redshift of $4.108 \pm 0.001$. The fainter, narrower component peaks at a redshift of $4.125\pm0.001$. If the narrow component is tracing the `systemic' velocity, then it appears that the line emission is dominated by the more turbulent, potentially outflowing component at the location of TNJ1338-b. As already mentioned in Section \ref{sec:regions}, both these components are significantly redshifted relative to TNJ1338-a. 

Outflow velocities can be estimated using the prescription from \citet{rupke05} using the \oiii\ lines. The outflow velocity, $v_\mathrm{out}$, is calculated using the peak velocity of the narrow component tracing the systemic emission ($v_\mathrm{n}$), the peak velocity of the broad component tracing the outflow ($v_\mathrm{b}$), and the velocity dispersion of the broad component ($\sigma_\mathrm{b}$). The $v_\mathrm{out}$ then follows from $|v_\mathrm{b} - v_\mathrm{n}| + 2\sigma_\mathrm{b}$.  Focusing on the results obtained for \oiii, which is the cleaner line complex of the two considered here, we find $v_\mathrm{out}=1400 \pm 40$\,\kms\ around TNJ1338-a and $1650 \pm 75$\,\kms\ for TNJ338-b.

\subsection{Dust attenuation, ionization state and metallicity}
\label{sec:ionization}

The 2D emission line intensity maps and the 1D spectra shown in the previous sections suggest that there is likely a complex range of ISM conditions across TNJ1338. In this section, we investigate the dust attenuation, ionization state and the metal distribution in the system.

We begin by exploring whether significant dust attenuation may be seen in the individual line emitting components. The typical approach would involve calculating the observed Balmer decrements (e.g. \ha/\hb) and comparing with theoretical predictions \citep[e.g., from][]{ost06} to calculate any observed reddening. However, as we have shown in the previous section, the relatively broad and kinematically disturbed \ha\ line appears to be blended with the \nii\ doublet, sometimes appearing to have multiple distinct components. Therefore, similarly high SNR would be needed in the \hb\ line to accurately derive the reddening of the individual line emitting components. 

Unfortunately for a part of the datacube, including TNJ1338-a, the \hb\ line falls in the detector chip gap. Therefore, as a first attempt to explore dust attenuation in the system, we focus on \hb/H$\gamma$ ratios in the regions where \hb\ is cleanly visible. For TNJ1338-a, we use the \ha/H$\gamma$ ratio to measure the observed Balmer decrement using the total \ha\ flux. We note here that the H$\gamma$ line appears to be blended with the \oiii\,$\lambda4363$ auroral line, so we perform a double Gaussian fit to disentangle the emission line fluxes. Overall, we find that across all the strong line emitting regions, the observed Balmer decrement ratios are very close to the theoretical expectations (i.e., H$\gamma$/\hb\ $\approx 0.47$), which implies little to no dust attenuation at the locations of the line emitting regions. A more detailed analysis of the 2D dust attenuation has been explored in a separate paper \citep{roy24}.

\begin{figure*}
    \centering
    \includegraphics[width=\linewidth]{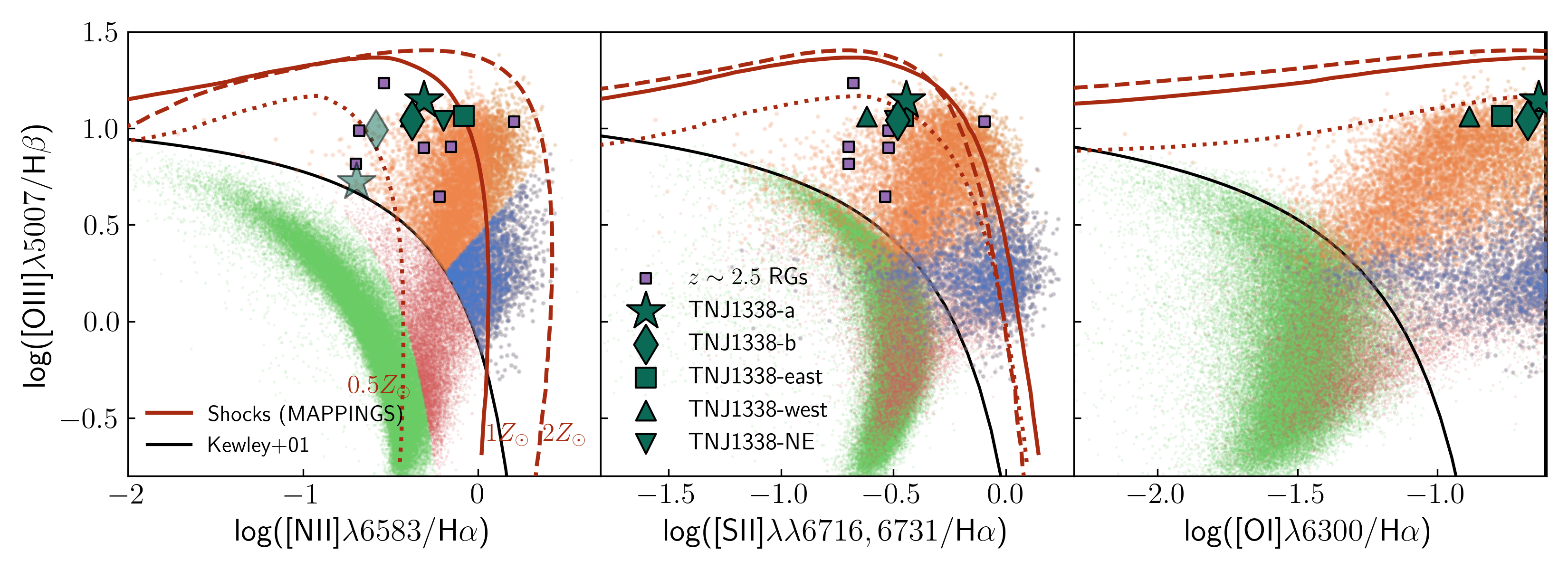}
    \caption{Location of the individual line emitting components identified in TNJ1338 on the \nii-BPT diagram (\citealt{baldwin81}, left), the \sii-VO87 diagram (\citealt{veilleux87}, middle) and the \oi/\ha\ vs \oiii/\hb\ diagram. Measurements from these components are shown as dark green symbols, where for TNJ1338-a and b we report two values, the solid marker is for the brighter component (narrow for TNJ1338-a and broad for TNJ1338-b) and the translucent marker is for the fainter line component. We also show data from SDSS, where green points are star-forming galaxies, orange points are AGN, red points are composite populations and blue points are LINERS. Measurements from $z\sim2.5$ radio galaxies are shown as purple squares, taken from \citet{humphrey08}. Also shown is the typical dividing line between star-formation and AGN/shocks derived by \citet{kewley06} in black, with predictions from photoionization due to shocks from \textsc{mappings} \citep{dopita96, allen08} in red, increasing in velocity from left to right. Very clearly, all the individual line emitting regions in this source lie away from the star-forming region of the diagnostic plot, and favour photoionization due to AGN and/or shocks. The shock models favor metallicities between $0.5-1\,Z_\odot$, with TNJ1338-b consistently requiring lower metallicities than TNJ1338-a.}
    \label{fig:BPT}
\end{figure*}

We now attempt to establish the dominant source of ionization in the line emitting regions, TNJ1338-a and TNJ1338-b. We employ the BPT \citep{baldwin81} and VO87 \citep{veilleux87} diagnostics that are based on the \nii\,$\lambda6583$/\ha, \sii\,$\lambda\lambda6716,6731$/\ha, and \oi\,$\lambda6300$/\ha\ versus \oiii\,$\lambda5007$/\hb\ ratios. These can be used to distinguish between star-forming regions and photoionization by AGN or shocks. As shown in Figures \ref{fig:tnj-a_linefits} and \ref{fig:tnj-b_linefits}, multiple component fits were used to model the observed \ha+\nii\ and \oiii\ line emission at the locations of TNJ1338-a and b. In this present study, we do not attempt to fit multiple components to the other regions, which will be explored in a future paper. We are further unable to conclusively obtain multi-component fits to the \sii\ and \oi\ emission lines, given their much lower SNR compared to the \ha+\nii\ complex. Therefore, only for regions TNJ1338-a and b, we measure the \nii-BPT line ratios individually for the two line components, as shown in Figure \ref{fig:BPT}.

In the \nii-BPT diagram, the filled star shows the location of TNJ1338-a for the strong (narrow) emission component with a higher total line flux, whereas the translucent star is the measurement from the weaker (broad) component in the \ha+\nii\ complex. The filled diamond is the measurement for the bright (broad) component of TNJ1338-b, and the translucent diamond is the weaker (narrow) component. For all other components and for the \sii-VO87 and \oi\ diagnostics, we only report the total line flux ratios. As can be seen from the figure, line ratios measured from both of the line components in TNJ1338-a and b, as well as at the locations of all other strong line emitting regions, lie away from the star-forming region of the BPT diagram, favouring photoionization driven by AGN/shocks as the dominant mechanism. 

The ratios measured from the fainter (broad) lines from TNJ1338-a lie closer to the star-forming part of the diagram compared to the brighter (narrow) emission. The line ratios shown in the figure are given in Table \ref{tab:line_ratios}. This analysis clearly demonstrates that none of the strong line emitting regions in TNJ1338 is tracing star-formation alone, and the bulk of the gas in this system is likely ionized by AGN or shocks. 
\begin{table*}
	\centering
	\caption{Emission line flux ratios used in the BPT and VO97 diagrams for the line-emitting components identified in TNJ1338. Note that for TNJ1338-a and b, we only give the \sii/\ha\ and \oi/\ha\ fluxes for the brighter of the two components fitted to the line complex.}
	\begin{tabular}{l c c c c}
		\toprule
		Component & \oiii$\lambda5007$/\hb\ & \nii$\lambda6583$/\ha\ & \sii$\lambda\lambda 6716,6731$/\ha\ & \oi$\lambda6300$/\ha\  \\
		\midrule
		TNJ1338-a (narrow) & $13.93 \pm 2.45^*$ & $0.49 \pm 0.10$ & $0.36 \pm 0.08$ & $0.23 \pm 0.05$ \\
		TNJ1338-a (broad) & $5.28 \pm 1.57^*$ & $0.20 \pm 0.08$ & $-$ & $-$ \\
		TNJ1338-b (narrow) & $9.82 \pm 0.78$ & $0.26 \pm 0.08$ & $-$ & $-$ \\
		TNJ1338-b (broad) & $11.03 \pm 0.10$ & $0.42 \pm 0.10$ & $0.33 \pm 0.03$ & $0.21 \pm 0.02$  \\
		TNJ1338-west & $11.55 \pm 0.78$ & $0.41 \pm 0.01$ & $0.24 \pm 0.02$ & $0.13 \pm 0.01$ \\
		TNJ1338-east & $11.59 \pm 1.41$ & $0.83 \pm 0.04$ & $0.35 \pm 0.13$ & $0.17 \pm 0.01$ \\ 
		TNJ1338-northwest & $8.08 \pm 0.75$ & $0.59 \pm 0.02$  & $0.32 \pm 0.12$ & $0.22 \pm 0.02$ \\ 
		\bottomrule
	\end{tabular}
	
	$^*$ \hb\ line measurement is affected due to \hb\ being in the chip gap. The \hb\ fluxes of the components have been estimated from the observed \ha\ fluxes, assuming no dust.
    \label{tab:line_ratios}
\end{table*}

In the figures we also show predictions from shock models from \textsc{mappings}\footnote{\url{https://mappings.readthedocs.io/en/latest/}} \citep{dopita96}, based on models of low-density, steady flow shocks with velocities in the range $150-500$\,\kms\ for three metallicities ($0.5\,Z_\odot$, $1\,Z_\odot$ and $2\,Z_\odot$, where $Z_\odot$ is solar metallicity). The purple squares show literature measurements for HzRGs at $z\sim2.5$ from \citet{humphrey08}, where the relevant lines could be observed from the ground. The line ratios that we measure from all strong line emitting components in TNJ1338 appear to overlap with the parameter space occupied by the lower redshift radio galaxies from the literature. Unfortunately, the predictions from shock models fully overlap with the AGN photoionization region in the BPT diagram, making it hard to conclusively differentiate between photoionization due to AGN or shocks based on this BPT analysis alone.

We do note here, however, that the distinction between the parameter space occupied by star-forming regions and AGN in the BPT diagram is expected to become less clear toward higher redshift \citep[e.g.,][]{groves06, kewley2013}. On the one hand, AGN may also occur in young star-forming galaxies that have relatively low metallicities compared to typical AGN host galaxies at low redshift. On the other hand, the low metallicity ISM of young star-forming galaxies can give rise to high ionization parameters and harder radiation fields from massive forming stars. Therefore, at $z\sim4$ it may not be straightforward to rule out photoionization driven by low-metallicity star-formation alone based on these diagnostics.

\begin{figure*}
    \centering
    \includegraphics[width=0.49\linewidth]{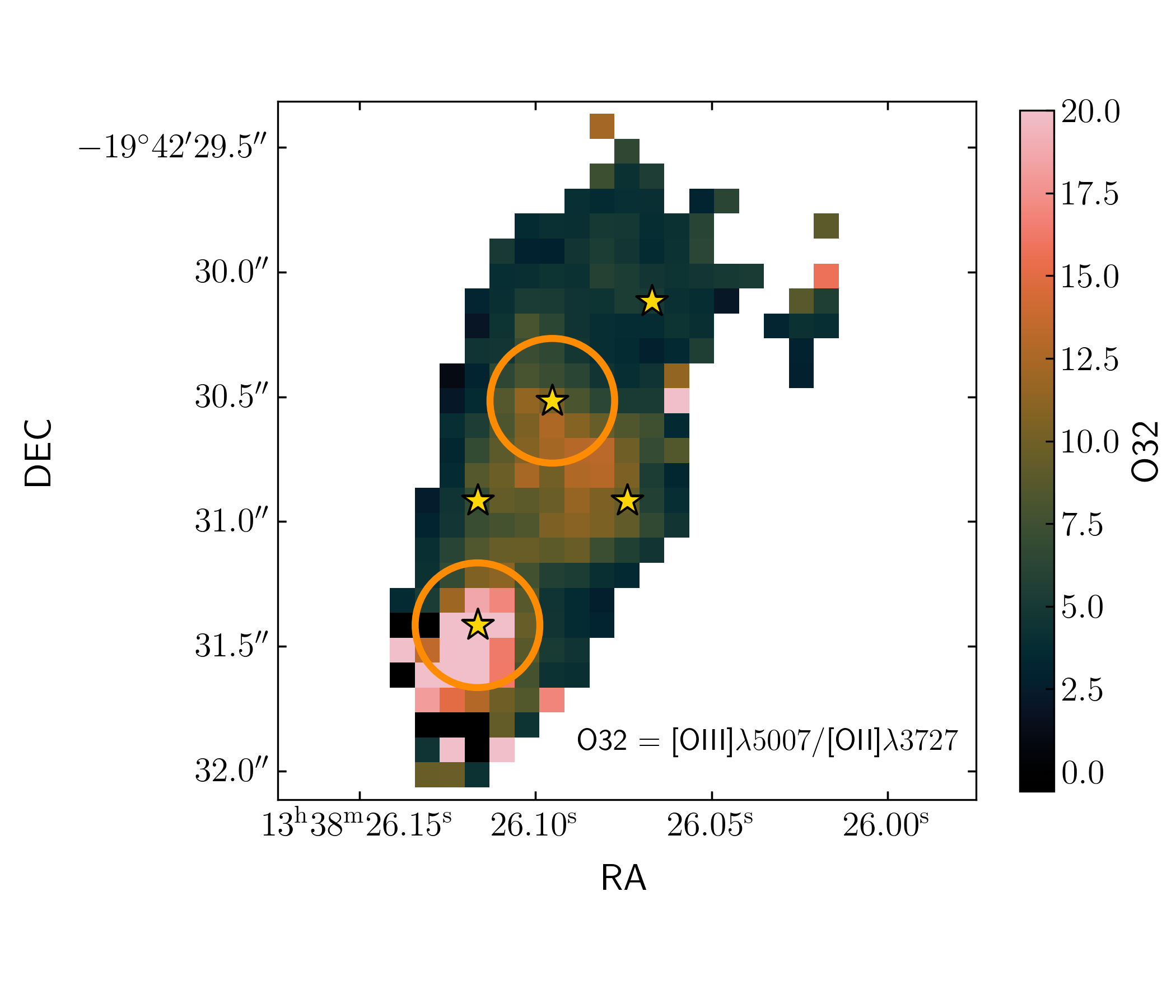}
   \includegraphics[width=0.49\linewidth]{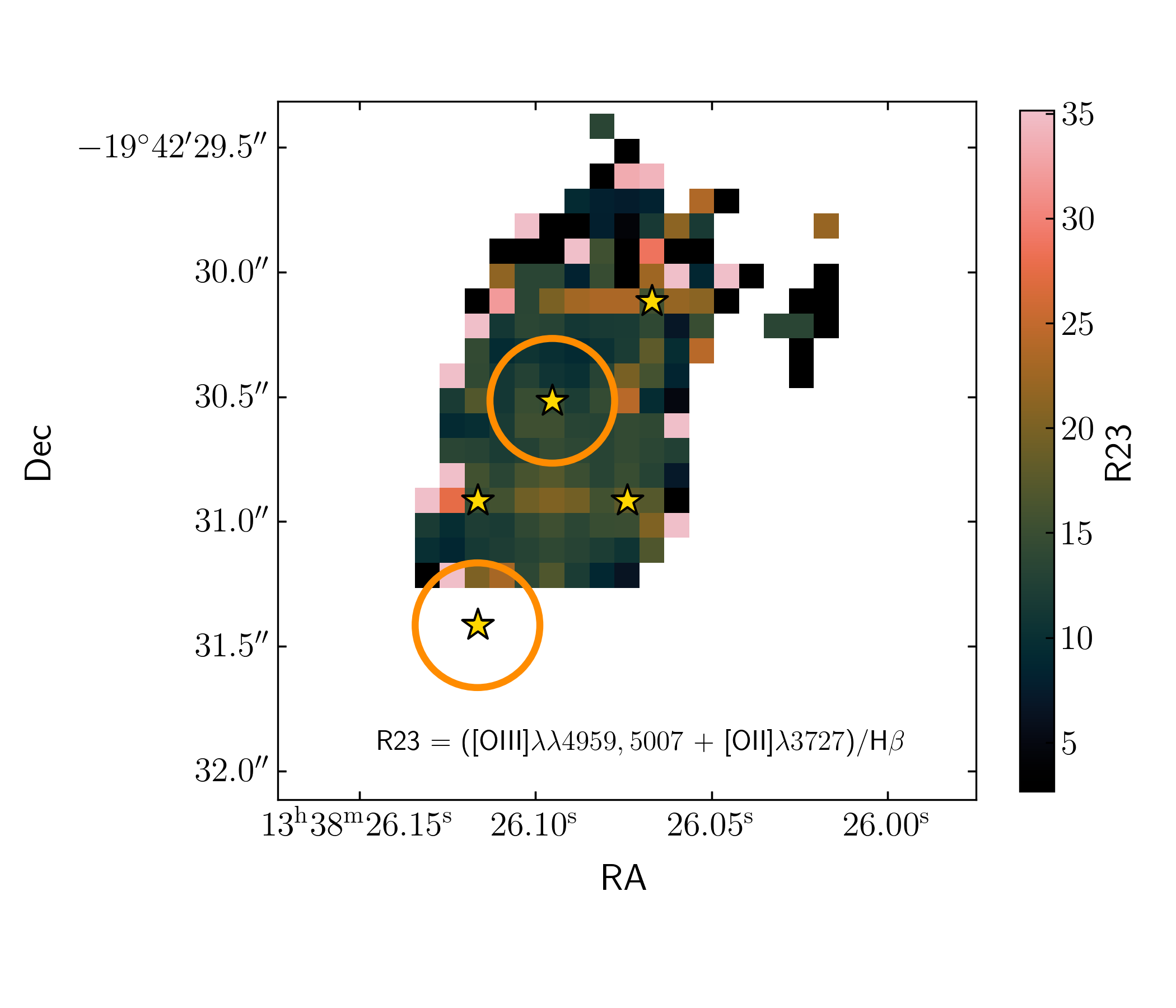}    
    \caption{2D map showing the spatial distribution of the \oiii$\lambda5007$/\oii$\lambda3727$ ratio (O32; left) and (\oiii$\lambda\lambda4959,5007$+\oii$\lambda3727$)/\hb\ (R23, right) in the system, with the apertures around TNJ1338-a and b marked using orange circles and the centroids of the individual line emitting regions are marked with gold stars. Unsurprisingly, TNJ1338-a shows some of the highest O32 ratios in the system. However, TNJ1338-b also traces a region of high O32 ratios, with a clear discontinuity visible in the distribution of O32 ratios between TNJ1338-a and b. The R23 map is affected by the \hb\ emission in the bottom half of the cube lying in the detector chip gap, which we have masked out here. Overall, the R23 distribution in the system appears relatively uniform, indicating a homogeneous distribution of metals across the system.}
    \label{fig:O32}
\end{figure*}

Further insights into the state of ionization can be gained by using the \oiii$\lambda5007$/\oii$\lambda3727$ ratio, or O32. O32 is sensitive to the ionization parameter and is independent of the gas-phase (O/H) abundance. In Figure \ref{fig:O32} we show the O32 2D map of the system, marking the locations of the centroids of the strong line emitting regions (gold stars). The highest O32 ratios are observed at the location of TNJ1338-a, highlighting perhaps unsurprisingly that the highest ionization parameters are seen around the main AGN in the system, with a secondary maximum near TNJ1338-b. 

Interestingly, the O32 ratios do not smoothly decrease with increasing distance away from TNJ1338-a, as they appear to do around TNJ1338-b. Low O32 ratios are observed in between these two regions, and then towards the northwest where the radio emission is the strongest. The fact that the O32 ratio map shows two clear local maxima is suggestive of completely separate ionizing sources at these two locations, however, it may also be possible that the two regions have very different physical conditions (like density, and temperature). Shock models employed specifically for bright radio galaxies in the literature have shown that shocked gas often emits relatively stronger lower ionization lines than AGN (or shocked precursor) photoionized gas \citep[e.g.,][]{clark1998, villar1999, best00}.

The western component (coinciding with the apex or tip of the \lya\ wedge, see Sect. \ref{sec:agn2}) has relatively high values of O32 as well, but interestingly the northwestern component does not, which can also be seen clearly in Figure \ref{fig:linemaps} where \oii\ emission appears to be bright near the northwest component. The northwestern component appears to be dominated by lower ionization emission.

The (\oii$\lambda3727$+\oiii$\lambda\lambda4959,5007$)/\hb\ (R23) ratio is a good tracer of the O/H distribution, regardless of the ionization parameter in the system, and this is shown in Figure \ref{fig:O32} (right). We note here again that \hb\ emission at the location of TNJ1338-a falls in the chip gap, and we are therefore unable to determine a reliable R23 ratio at this location. In contrast to the O32 ratio, we do not note an appreciable change in the R23 value across the system. Therefore, the O32 and R23 ratios together seem to indicate that although the ionization parameter across the system shows a lot of structure, the metallicity distribution remains roughly the same across all strong line emitting regions. This is consistent with what was reported for AGN selected from SDSS by \citet{dors2021}, where no R23 variations were reported along the radii of the AGN. \citet{dors2021} also found a positive correlation between R23 and O/H in AGNs, indicating that the hardness of the ionization field is not affected by the metallicities in these objects, similar to what is revealed by the O32 and R23 ratios in TNJ1338. 

\subsection{High-ionization emission lines}
\label{sec:nev}

Further insights on the extreme ionization conditions in this system can be gained based on the presence of certain high ionization emission lines. \citet{nesvadba17a} found evidence of extended \neiii\,$\lambda3869$ in TNJ1338, which can be clearly seen centered around the locations of TNJ1338-a and TNJ1338-b in the line intensity map shown in Figure \ref{fig:linemaps}. Our 1D spectrum at the location of TNJ1338-a also shows significant \nev\ emission, which requires an extremely high ionization energy $E > 97.1$\,eV, and is therefore a tracer of hard photoionizing spectra of (obscured) AGN \citep[e.g.][]{gilli10,mignoli13}. It can also trace fast shocks in AGN  \citep[e.g.][]{leung21,cleri23} or star-forming regions due to highly energetic supernova shocks \citep[e.g.][]{izotov12}. We further note the presence of \heii\,$\lambda4686$ emission at the locations of TNJ1338-a and b, which also requires high energy photons with $E>54.5$\,eV.

\begin{figure}
    \centering
    \includegraphics[width=\linewidth]{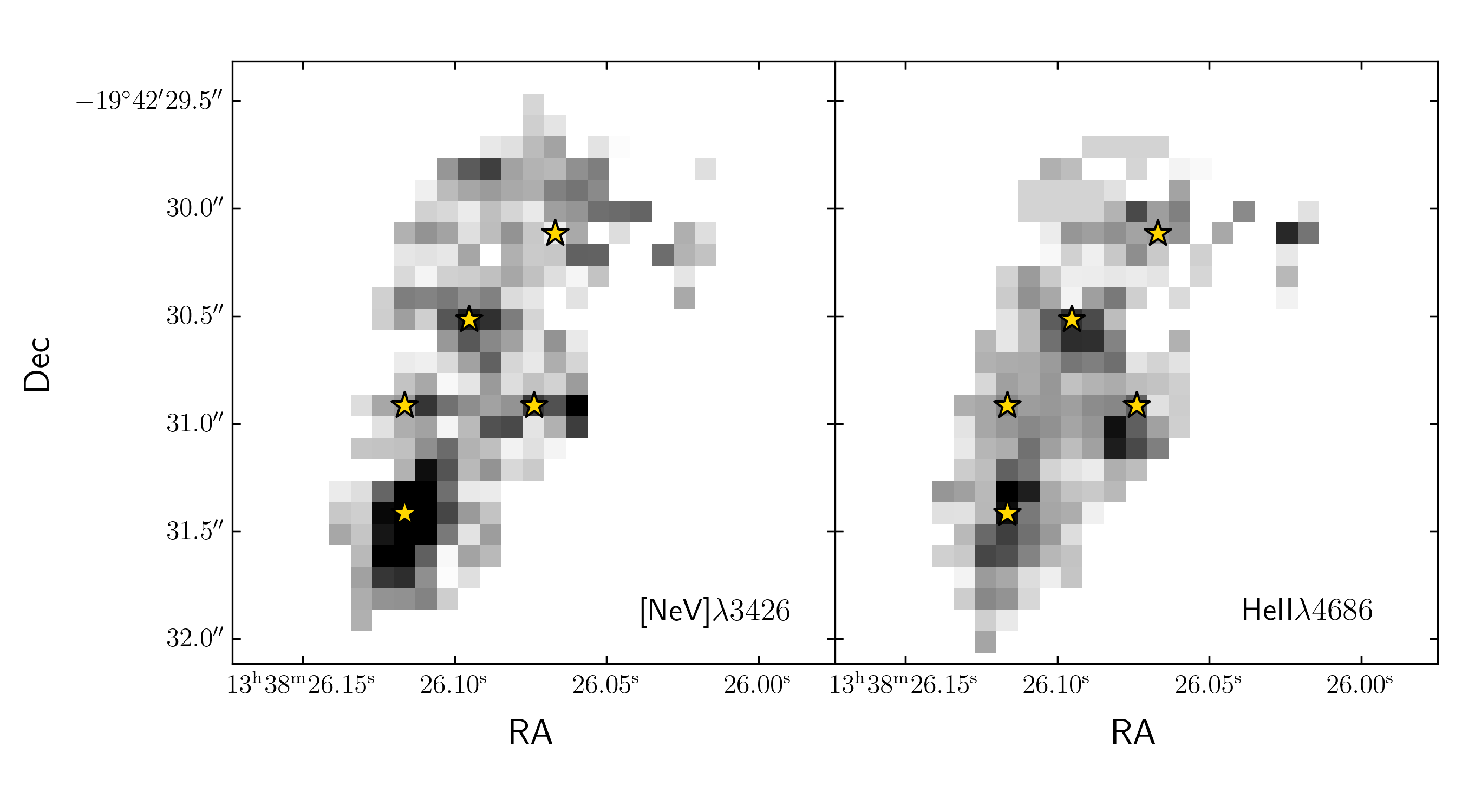}
    \caption{\nev$\lambda3426$ (left) and \heii$\lambda4686$ (right) 2D maps, tracing some of the highest ionization energy lines in the system, with the locations of the apertures from Fig. \ref{fig:TNJ-overview} marked (stars). \nev\ emission from TNJ1338-a is clearly detected, while \nev\ emission from TNJ1338-b is significant at the $\sim2\sigma$ level. Since the peak of this emission coincides with the peak emission from other lines at TNJ1338-b, the \nev\ emission that we see is likely to be real and not due to a significant noise fluctuation. Shown right, \heii\ emission from both TNJ1338-a and b is clearly detected, with strong emission also seen coinciding with TNJ1338-west, towards the wedge. The upper part of the \heii\ emission map begins overlapping with the chip gap, but this does not affect a large part of the map from where the main strong line emission is coming.}
    \label{fig:nev}
\end{figure}

To investigate the nature of this high ionization energy emission, in Figure \ref{fig:nev} we show the continuum-subtracted line map of \nev\,$\lambda3426$ (left panel) and \heii\,$\lambda4686$ (right panel). Qualitatively speaking, component TNJ1338-a is clearly associated with strong \nev\ and \heii\ emission. Interestingly, a point-like source of \nev\ emission is also detected at the location of TNJ1338-b, although at much lower significance ($\sim2\sigma$). The \heii\ emission at this location is much clearer. The fact that the centroid of \nev\ emission at TNJ1338-b coincides perfectly with the location of \neiii, 
\heii\ and other much stronger lines (Figure \ref{fig:linemaps}) increases the likelihood that this feature is real and not due to noise fluctuations. We further note the possible faint detection of \nev\ and \heii\ at the location of the western component as well.

Measuring fluxes directly from the continuum-subtracted 2D images in the same apertures across TNJ1338-a and b, we infer \nev/\neiii\ (Ne53) ratios of $0.68 \pm 0.01$ and $0.28 \pm 0.10$, respectively. Combined with O32 ratios for TNJ1338-a and b of $13.71 \pm 0.01$ and $6.93 \pm 0.02$ and the diagnostic plot presented in \citet{leung21}, which employ predictions from faster shock models implemented by \citet{allen08} with maximum shock velocities going up to 1000\,\kms, this places TNJ338-a just outside the region affected by shock-ionization, making AGN photoionization a more likely choice. The results for TNJ1338-b are inconclusive because it falls in a region of the O32-Ne53 space in which photoionization and shock-models overlap. TNJ1338-b could be explained by either standard AGN photoionization or by (very) fast ($v>600$\,\kms) shocks. 

Using the shock models from \citet{dopita96} with a maximum velocity of 500\,\kms, \citet{best00} explored a similar line ratio diagnostic employing Ne53 and the semi-forbidden C transitions C\,\textsc{iii}]$\lambda1909$/C\,\textsc{ii}]$\lambda2326$ (C32). Although the rest-UV C lines are not covered by our spectra, one may substitute the C32 ratio with the O32 ratio assuming that the C/O abundance ratio does not change dramatically across the system (as evidenced by the O/H distribution traced by R23). The C\,\textsc{iii}] line has an ionization potential of $24.4$\,eV, whereas the C\,\textsc{ii}] has an ionization potential of $11.3$\,eV. By comparison, \oiii\ has an ionization potential of $35.1$\,eV and \oii\ has $13.6$\,eV. Since the ionization potentials of these lines are comparable, the O32 ratio may serve as a rough proxy for the C32 ratio. Using this argument, the \citet{best00} models would classify both TNJ1338-a and TNJ1338-b as being photoionized by an AGN rather than shocks. 

The inference made using the line ratio diagnostic plots from \citet{best00} and \citet{leung21} imply that shocks with $v<500$\,\kms\ are unlikely to explain the \nev\ and \neiii\ strengths seen at the location of TNJ1338-b, and would favour photoionization by AGN. Faster shocks, however, may explain the observed line ratios, but these predictions also completely overlap with the parameter space occupied by AGN models. Therefore, using the \nev\ and \neiii\ emission lines we cannot conclusively rule out photoionization by an AGN at the location of TNJ1338-b.

The \heii\ detection also tells a similar story. At the location of TNJ1338-b, where the \hb\ line is cleanly detected, we measure a \heii/\hb\ ratio of $0.095 \pm 0.010$, which is consistent with photoionization due to AGN following the line ratio diagnostics developed by \citet{baer2017}, where they showed that the \heii-based diagnostics were much more efficient at identifying AGN compared to traditional BPT diagnostics. As noted earlier, \heii\ emission is also seen at the location of TNJ1338-west, which we will explore in detail in a future paper. We will return to the possible nature of TNJ1338-b in Section \ref{sec:agn2}. In Table \ref{tab:ion_ratios} we list the measured line ratios tracing the ionization conditions at the locations of TNJ1338-a and b.
\begin{table}
    \centering
    \caption{Emission line ratios tracing the ionization conditions at the locations of TNJ1338-a and b.}
    \begin{tabular}{l c c c c}
        \toprule
         & O32 & R23 & \nev/\neiii\ & \heii/\hb \\
        \midrule
        TNJ-a & $13.71 \pm 0.01$ & $-$ & $0.68 \pm 0.01$ & $-$ \\
        TNJ-b & $6.93 \pm 0.02$ & $14.54 \pm 0.50$ & $0.28 \pm 0.10$ & $0.095 \pm 0.010$ \\
        \bottomrule
    \end{tabular}
    \label{tab:ion_ratios}
\end{table}

\subsection{Large scale kinematics}

\begin{figure*}
    \centering
    \includegraphics[width=\linewidth]{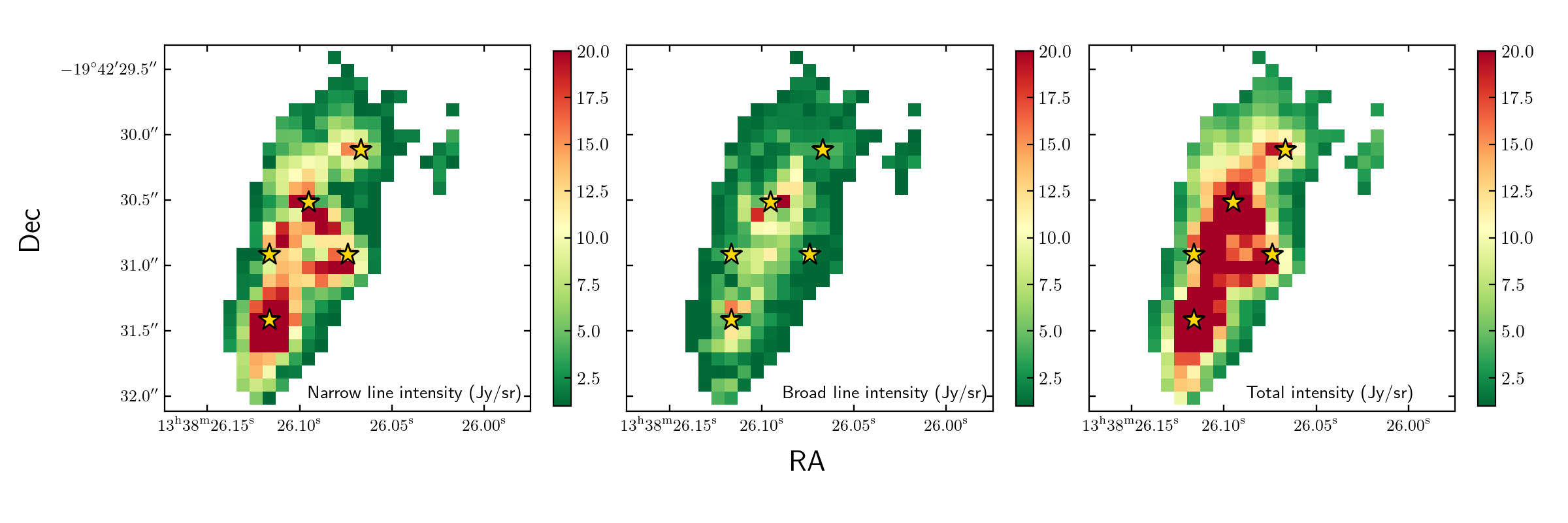}
    \caption{\oiii\ line intensity (right) separated into the best-fitting narrow (left) and broad (middle) emission components, with the total line intensity shown in the right panel. This figure illustrates that the broad components of \oiii\ emission in TNJ1338 seem to align along the radio jet axis, showing the turbulence (and ionization) injected by the radio gas in the system. The narrow line emission is strongest at the location of TNJ1338-a, and TNJ1338-b also shows a good mix of narrow and broad \oiii\ components.}
    \label{fig:oiii-intensities}
\end{figure*}

\begin{figure*}
    \centering
    \includegraphics[width=\linewidth]{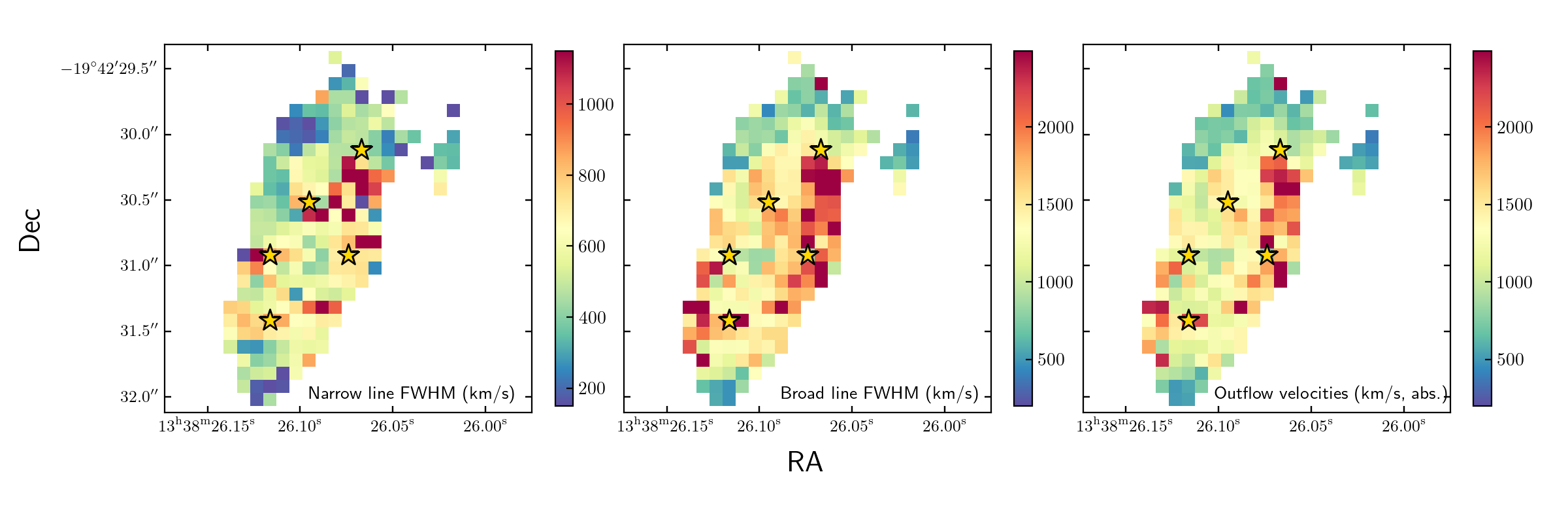}
    \caption{FWHM of the narrow (left) and broad (middle) components of \oiii, along with the outflow velocities inferred using the \citet{rupke05} prescription (right). We note that both the FWHM of the narrow and broad components in the system appear to peak towards the eastern/northeastern edge, close to where the \lya\ wedge and the radio hotspot were also reported. The outflow velocities are clearly highest around this region as well, which are calculated mainly using the velocity shift between the narrow and broad line component. This clearly shows increased turbulence in the gas closer to the radio hotspot.}
    \label{fig:oiii-velocities}
\end{figure*}

In Section \ref{sec:kinematics} we showed the evidence for strong narrow and broad velocity components associated with the apertures centered on TNJ1338-a and TNJ1338-b. Here we will apply this analysis to the entire system on a spaxel-by-spaxel basis, focusing on the strong \oiii\ emission line. To do this, we extracted a sub-cube around the \oiii\ line and then fit our two-component Gaussian model to the line in every spaxel exactly as in Section \ref{sec:kinematics}. We then plot the line intensities and velocities for the narrow and broad components of the best-fitting model separately. We do not require a minimum value of FWHM for the component to be classified as broad -- we simply label the broader of the two components to as broad. Figures \ref{fig:oiii-intensities} and \ref{fig:oiii-velocities} show the \oiii\ line intensity and velocity maps, respectively.

As discussed before and as shown in Figure \ref{fig:oiii-intensities}, the narrow \oiii\ emission is most prominent at the location of TNJ1338-a. There is appreciable narrow \oiii\ emission also seen at the location of TNJ1338-b, as well as in the west and east components. The broad \oiii\ line flux map presents a very different picture, dominated by an almost continuous linear feature extending from TNJ1338-a to roughly the location of the VLA radio emission at its Northern tip. We note that this is somewhat towards the west of the location of TNJ1338-b, which is prominently seen in the narrow line intensity map. As such, this linear broad line intensity feature appears to follow the expected direction of the hidden radio jet. An alternative explanation is that the broad line flux is somehow associated with TNJ1338-b, which will be explored as part of an alternative scenario given in Sect. \ref{sec:agn2}.     

Figure \ref{fig:oiii-velocities} shows the spatial distributions of the FWHMs of the narrow and broad components of \oiii. Besides the fact that there are almost no spaxels where the velocity drops below $\sim$500\,\kms, the most striking feature is a region on the western edge of the galaxy where both the narrow and the broad components seem to reach their maximum velocities (about 1000 and 2000\,\kms, respectively). It is interesting to note that this region does not coincide with any particular high intensity emission line feature (see right panel of Figure \ref{fig:oiii-intensities}). Curiously, it lies even to the west of the radio hotspot traced by the VLA radio emission. Following the outflow calculation of Section \ref{sec:kinematics}, this region naturally coincides with the region having the highest outflow velocities of around 2000\,\kms\ (right panel of Figure \ref{fig:oiii-velocities}). 

\section{On the nature of TNJ1338}
\label{sec:discussion}

\subsection{The AGN nucleus: TNJ1338-a}

In the initial discovery paper \citep{debreuck99}, a very faint, steep-spectrum radio component seen close to the Northern hotspot in the maps from \citet{pentericci00} was interpreted as the radio core. Registration to a ground-based $K$-band image placed this component just South of the host galaxy, with the Northern hotspot falling near the peak of the host galaxy in the ground-based image. However, this registration suffered from the $\sim$0.4\arcsec\ uncertainty in the optical reference system \citep{debreuck99}. High resolution images with the \emph{HST}/ACS later showed that the presumed faint radio core was more likely to be associated with the peak intensities in the rest-UV and $K$-band images with a relative uncertainty of 0.3\arcsec\ \citep{miley04,zirm05}. 

This also shifted the Northern hotspot to lie further to the northwest of the galaxy. This registration was substantially improved by \citet{duncan23} who achieved a $\sim2$\,mas accuracy astrometry tied to GAIA DR3. That registration was also adopted here and in \citet{roy24}, and should settle the matter at least where it concerns the location of the radio hotspot (see Figure \ref{fig:TNJ-overview}). However, our own analysis of the faint radio feature previously associated with the core of the AGN cannot confirm if this radio component is real or not. Without the definite identification of a radio core location, it remained to be shown from where the radio jets in TNJ1338 are launched. 

Based on the analysis presented in this paper, there is now ample evidence that TNJ1338-a is indeed the location of the SMBH powering the radio emission. In particular, the results from the BPT analysis, the very high O32 ratio and the high ionization \nev\ emission point to a strong source of photoionization at that location. The lack of BLR emission found in Figure \ref{fig:tnj-a_linefits} further confirms it to be an obscured (Type 2) AGN. This component also dominates the total and narrow-line \oiii\ intensity, and is associated with significant outflow velocities of $\sim$1500-2000\,\kms. TNJ1338-a furthermore coincides with the region where the highest stellar mass and the highest stellar mass surface density are found based on NIRCam images \citep{duncan23}, making it the most likely site for a SMBH in this entire system. As shown in Sect. \ref{sec:results}, the redshift of this component is $z=4.102$.

In \citet{roy24}, a more detailed analysis of the kinematics of TNJ1338 is presented. Comparison of the total ionized gas mass and mass outflow rates with the available kinetic energy input from the radio jet implies that the jet couples only weakly to the ISM. The overall picture is consistent with the standard scenario for an expanding radio cocoon \citep{begelman89}.  

\subsection{Evidence for a dual-AGN system?}
\label{sec:agn2}

The previous section shows that the region coinciding with component TNJ1338-a has been identified as the most likely location for the SMBH that is powering the radio jets. In a separate paper, it is shown that photoionization by the AGN located at TNJ1338-a is, in principle, sufficient for providing the source of ionization to scales of about 10 kpc along a radiation cone aligned with the radio axis \citep{roy24}. Curiously, however, as shown in this paper there are strong observational similarities between TNJ1338-a and TNJ1338-b. The latter component was identified based on having the strongest line emission outside the nucleus (Figures \ref{fig:TNJ-overview} and \ref{fig:oiii-intensities}). Similar to TNJ1338-a, it has a spectrum consistent with AGN photoionization (Figure \ref{fig:BPT}) and a similar \oiii\ line profile consisting of a narrow and broad component (Figure \ref{fig:tnj-b_linefits}), albeit peaking at different velocities. The fact that the same broad velocity component was seen both for \ha\ and \oiii\ makes it unlikely that we are seeing a BLR, just as in the case of TNJ1338-a. 

TNJ1338-b also shows evidence for a \nev\ point source, albeit with fainter flux compared to that seen at the location of TNJ1338-a (Figure \ref{fig:nev}). If we were to apply to the same reasoning to TNJ1338-b that was used to identify TNJ1338-a with an (obscured) AGN, we would reach the conclusion that TNJ1338-b is also an obscured AGN. If this interpretation is correct, this would be the first dual AGN system discovered in a high redshift radio galaxy with potentially very interesting implications. However, before reaching such a conclusion, an important complicating factor needs to be taken into account. As can be seen in Figure \ref{fig:TNJ-overview}, location TNJ1338-b lies quite close to the location of the Northern hotspot. It is therefore important to investigate whether the high ionization, the large line velocities, and the \nev\ component are not somehow related to the radio feature. 

Although the TNJ1338-b component lies close to the Northern hotspot, they are not exactly spatially coincident. As can be seen in Figure \ref{fig:TNJ-overview}, the TNJ1338-b centroid lies about $0.2-0.3$\arcsec\ SE of the hotspot centroid. TNJ1338-b also does not lie on the connecting line between TNJ1338-a (now identified with the core of the radio galaxy) and the radio hotspot. These offsets are much larger than the combined uncertainties in the astrometric solution of the NIRCam (and hence NIRSpec) and VLA images ($\sim$2 mas). 

As ionization of Ne$^{3+}$ requires soft X-ray photons with energies in the range 97-126 eV, the \nev\ $\lambda\lambda$3346,3426\AA\ emission line doublet is a common tracer of (obscured) AGN \citep[e.g.][]{schmidt98,gilli10,cleri23}. TNJ1338-b has  $\log(L_\mathrm{\nev}/\textrm{erg/s})= 43.5 \pm 0.3$, which is comparable to typical Type 2 AGN at lower redshifts \citep[e.g.,][]{gilli10}. However, as shown in Section \ref{sec:nev}, diagnostic diagrams involving the Ne53 line ratio \citep{best00,leung21} do not provide a conclusive answer on whether TNJ1338-b is powered by high velocity shocks associated with the impacting radio jet or by another photoionization source. Figure \ref{fig:O32} shows a secondary peak in O32. This peak does not coincide with the hotspot location, but is located in between TNJ1338-b and component `west'. If there is a secondary AGN photoionizing the ISM, we would expect O32 to decrease with increasing distance from this particular region, which is consistent with what is observed (Figure \ref{fig:O32}). 

\begin{figure*}
    \centering
    \includegraphics[width=0.8\linewidth]{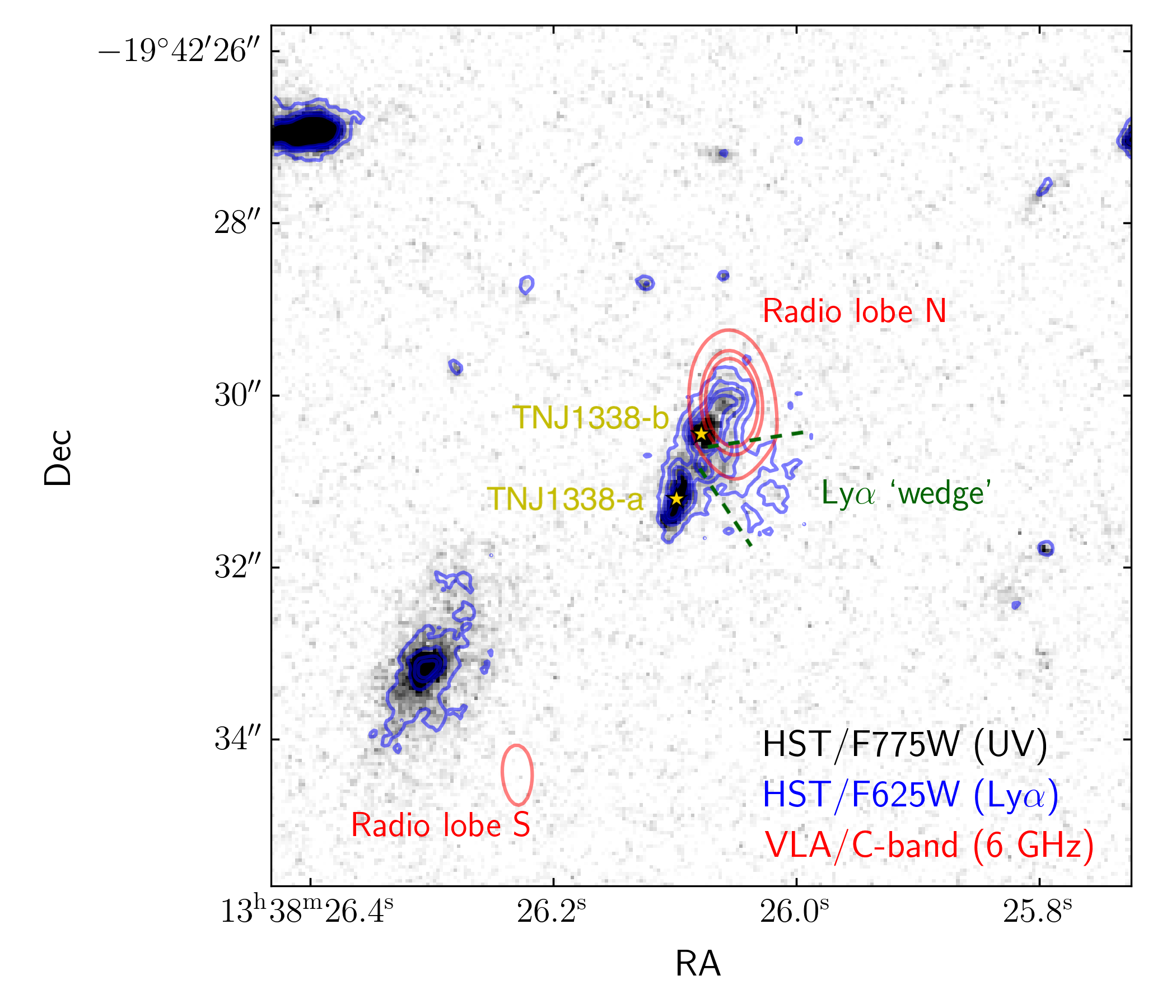}
    \caption{\emph{HST}/ACS F775W image from \citet{miley04}, along with contours from F625W that contains prominent and extended \lya\ emission (blue) including the `wedge' feature discussed in Section \ref{sec:agn2} and \citet{zirm05}. We also show VLA C-band (6\,GHz) contours from \citet{pentericci99}, which shows the brighter northern and the fainter southern radio lobes. The wedge is suggestive of a second ionization cone that is oriented along the direction to TNJ1338-b (dashed line). This direction is perpendicular to the radio axis that goes through TNJ1338-a.}
    \label{fig:wedge}
\end{figure*}

A very interesting feature of TNJ1338 that was highlighted by \citet{zirm05} is a wedge-shaped region of \lya. The wedge has a rest-frame \lya\ EW of $>$650 \AA\ and is easily detected in the ACS $r_{625}$ image (Figure \ref{fig:wedge}; see also Figure 4 in \citet{duncan23}). While HzRGs frequently show line and continuum emission that is aligned with their radio axes, it is rare to find strong features that are perpendicular to it. 
\citet{zirm05} performed an extensive analysis of the wedge. Its apex lies near component `west' and it extends for about 15 kpc toward the west away from the main body of the galaxy defined by the continuum and line emission. Radial and azimuthal surface brightness profiles indicate that it has a relatively sharp edge at 10 kpc and reaches maximum brightness along the direction that is perpendicular to the radio axis with relatively sharp cut-offs at either side of the azimuthal profile. There is some evidence for a much fainter counterwedge on the opposite side of the galaxy. With its apex about 1\arcsec\ away from TNJ1338-a and its main axis furthermore perpendicular to the radio axis, \citet{zirm05} argued that it is unlikely that the wedge is due to photoionization by the AGN or related to shocks from the radio jets. They furthermore ruled out star-formation as a source of photoionization within the wedge itself, based on the very high EW. 

Instead, they interpreted the wedge as evidence for a supernova-driven outflow in which the gas is photoionized at its apex and shock-ionized further out, similar to that observed in local starburst galaxies \citep{heckman90}. They also showed that the wedge naturally connects to the larger scale \lya\ halo that extends over 150 kpc, roughly aligned with the radio axis in its outskirts (see their Figure 11). A possible scenario is thus that the wedge traces outflowing gas that eventually streams along the boundary of the cavity excavated by the radio jets. 

\citet{duncan23} searched for a counterpart of the \lya\ wedge in the NIRCam F335M image that contains \ha+\nii, but found none. 
However, we found that the apex of the wedge coincides with the distinct feature that we have labeled 'west' in Figure \ref{fig:TNJ-overview} (it can also be seen in the NIRCam images). Although there is no obvious counterpart of the further wedge in the rest-frame optical emission line maps (Figure \ref{fig:linemaps}), a faint, isolated feature that can be seen in the top-right in all line maps does coincide with a similar feature in \lya\ seen just at the edge of the wedge.

Interestingly, comparing Figs. \ref{fig:O32} and \ref{fig:oiii-velocities} shows that the O32 peak near TNJ1338-b does not coincide with the region where the highest outflow velocities are observed. However, the latter does coincide with the apex of the wedge. In the dual AGN scenario, these high velocities could perhaps originate from strong radiative feedback from the second AGN. Given the connection between the wedge and the larger-scale \lya\ halo, this outflow could then ultimately even be responsible for feeding the \lya\ halo.   

Further, \citet{zirm05} concluded that if a second obscured (and radio-quiet) AGN would be located at the component that we now identify as TNJ1338-b, this would explain the wedge feature tracing a photoionization cone arising from this AGN. They concluded that while not impossible, this explanation was not preferred mainly given the lack of significant continuum in the $K_S$-band at that location. We have shown that components TNJ1338-a and TNJ1338-b have a systemic velocity difference of about 411\,\kms. If TNJ1338-b also hosts a SMBH, then one possibility is that it is part of a companion galaxy that is merging with TNJ1338-a, given the large velocity difference. Figure \ref{fig:linemaps} and the analysis in \citet{duncan23} show that although there is no obvious secondary galaxy at this location, there is a significant amount of rather diffuse stellar continuum. Re-analysis of the stellar mass density maps of \citet{duncan23} incorporating the additional continuum information provided by the NIRSpec/IFU observations indicates a (maximum) stellar mass of $\log(M_\star/M_\odot)\approx 9.8$ inside the aperture of TNJ1338-b, roughly a factor 10 less than that measured for TNJ1338-a. However, this mass estimate is similar to that found in other, randomly placed apertures around this area, indicating that diffuse stellar mass associated with for example tidal debris as opposed to the bulk mass of a merging companion is another possibility.

In light of the evidence we have presented for a possible second AGN associated with TNJ1338-b, a conclusion that was also independently derived by \citet{zirm05} purely based on the location and morphology of the wedge feature, we conclude that the possibility of a second AGN in TNJ1338 is still very much an option. 

\subsection{Implications for the host galaxy}
The line widths and the outflow velocity we infer at the location of TNJ1338-a are comparable to what has been previously observed for bright radio galaxies and X-ray selected AGN at $z\sim2$ \citep[e.g.][]{nesvadba17a,kakkad20}. These findings are also consistent with what is observed in young radio galaxies at $z<1$ \citep{santoro20}. Compared to outflows from more `normal' high-mass galaxies with no obvious presence of a bright AGN at $3<z<9$ \citep[e.g.][]{carniani23}, the line widths and outflow velocities in TNJ1338 appear to be much higher, suggesting a strong role played by the AGN in the broadening of the lines and resulting in a galaxy scale outflow. A detailed study of these large-scale outflow effects is presented in \citet{roy24}.  

The results in this paper paint a somewhat different picture than that presented in \citet{zirm05} and \citet{duncan23}, who found the photometric observations of the line emission in TNJ1338 to be consistent with an ISM, possibly triggered by positive AGN feedback from the large scale radio jets. Instead, the NIRSpec data show that the ISM is almost entirely photoionized by AGN with some contribution from shocks and none of the individual components is consistent with photo-ionization from star-formation alone. However, the stellar mass and mass surface density estimates derived by \citet{duncan23} should remain valid, given that this is largely determined by the emission line free NIRCam filters. 

Furthermore, the value of $M_\star\sim10^{11}$ $M_\odot$ derived by \citet{duncan23} is consistent with previous estimates based on lower resolution data from \emph{HST} and \emph{Spitzer} \citep{overzier09,debreuck10}. One concern was that if TNJ1338 included a significant amount of scattered continuum or broad line region emission from an AGN, this could have significantly affected these stellar mass estimates. The fact that we do not find evidence for any BLR means that the high stellar mass determined by \citet{duncan23} likely remains largely valid. 
 
We can use the observed \oiii\ luminosity to estimate the bolometric luminosity of the AGN using the result from \citet{heckman04}, who found $L_\mathrm{bol}=3500\times L_\mathrm{\oiii}$ erg\,s$^{-1}$ with a dispersion of 0.38 dex. Here $L_\mathrm{\oiii}$ is based on the observed \oiii$\lambda5007$ flux not corrected for dust. Taking only the total line luminosity directly associated with TNJ1338-a we find $L_\mathrm{bol}=5.0\times10^{48}$ erg s$^{-1}$. The mass of the black hole can then be calculated as \citep{jia11} $M_\bullet=2.6\times10^{-5}L_\mathrm{bol}(L_\odot)(L/L_\mathrm{E})^{-1}=3.4\times10^{10}(L/L_\mathrm{E})^{-1}\,M_\odot$, where $(L/L_\mathrm{E})$ is the Eddington ratio. 

Although we have no observational probe of the actual Eddington ratio, any value less than unity would quickly result in a SMBH of unphysical proportions \citep{mcconnell11}. The large (inferred) black hole mass in our calculation is mainly driven by the very high $L_\mathrm{\oiii}$. However, it must be noted that the line emission in TNJ1338 is extremely luminous compared to typical HzRGs. Its \ha\ luminosity, for example, is about 10 times more luminous than the median value found in the sample of HzRGs studied by \citet{nesvadba17a}. A contribution from star formation is unlikely to be the cause of this excess, as the BPT analysis clearly shows that AGN-related processes are dominant. Assuming no dust and taking the SFR of $\sim$500 $M_\odot$ yr$^{-1}$ (see below) would contribute about $8\times10^{43}$ erg s$^{-1}$ to the \oiii$\lambda5007$ luminosity \citep{villa-velez21}, which is about 5\% of the total observed. 

An alternative estimate of the bolometric luminosity, derived from the total IR luminosity of the best-fit AGN model from \citet{falkendal19}, is $2\times10^{47}$ erg s$^{-1}$. Using this value instead, we find $M_\bullet=1.4\times10^{9}(L/L_\mathrm{E})^{-1}\,M_\odot$. This is more than 10$\times$ lower than the above estimate that was based on $L_\mathrm{\oiii}$, and furthermore a value that is much more comfortable for a wider range of Eddington ratios. In the dual AGN scenario, if we assume about half of $L_\mathrm{bol}$ for each of the AGN, the individual BHs would each have masses of order $10^{9}(L/L_\mathrm{E})^{-1}\,M_\odot$. 

Together with the stellar mass estimates for the two components, and assuming Eddington accretion, this would imply black hole mass to stellar mass ratios of order $0.01$ and $0.1$ for TNJ1338-a and TNJ1338-b, respectively. This difference is entirely driven by the factor 10 difference in their stellar masses. While the value for TNJ1338-a is reasonable for luminous AGN at high redshift \citep[e.g.][]{neeleman21}, the value for TNJ1338-b is very much higher. The latter would be consistent with the possible scenario scenario of a SMBH found amidst tidal debris. We note here that recent \emph{JWST} observations of AGN at $z>4$ selected via the presence of broad Balmer line emission have also showed similarly high $M_{\rm{BH}}/M_{\star}$ ratios \citep[e.g.][]{harikane23, maiolino23}, which lie above the locally observed scaling relations and may be indicative of an earlier assembly epoch for SMBHs, and a later epoch of stellar mass assembly. \citet{maiolino23} noted that the BH masses and velocity dispersion are more in line with local relations, which for TNJ1338 we will explore further in a future study.

Despite our new interpretation of the dominant ionization mechanism powering the UV continuum and optical line emission being AGN rather than star-formation, TNJ1338 is likely not a quiescent galaxy. TNJ1338 was also detected at 450 and 850 $\mu$m and 1.2 mm \citep{debreuck04}, and the SED was modeled by \citet{falkendal19} who derived a total SFR of 461 $M_\odot$ yr$^{-1}$. These wavelengths are in between the power-law and warm dust emission associated with AGN at mid-IR wavelengths and the synchrotron emission at radio wavelengths. In that particular region, the spectrum is entirely dominated by the modified black body spectrum expected from cold dust heated by star formation. \citet{roy24} have modelled the global kinematics of TNJ1338, and find a total mass outflow rate of about 500\,$M_\odot$ yr$^{-1}$. This is roughly equal to the far-IR/sub-mm SFR. With a mass loading factor of order unity, the galaxy thus appears to be heavily limited by the kinetic feedback associated with the radio jets.  

\subsection{Implications for the protocluster environment}

TNJ1338 has received significant attention since its discovery, because of evidence that it is located near the center of a large overdensity of galaxies  \citep{venemans02,miley04,intema06,venemans07,overzier08,overzier09,debreuck10,saito15}. Analysis of the size and magnitude of the overdensity has shown that the TNJ1338 structure is consistent with the progenitor of a rich cluster of galaxies that will collapse and virialize some time before the present day. The massive host galaxy and the powerful AGN-driven feedback observed make TNJ1338 a clear candidate for what may become the brightest cluster galaxy in the future cluster.   

\citet{smail13} detected extended X-ray emission with an (observed) 0.5-8 keV luminosity of about $2.2\times10^{44}$ erg s$^{-1}$. The emission extends over 30 kpc roughly aligned with the radio source. This X-ray emission is much too high for the thermal Bremsstrahlung expected from any virialized system at $z\sim4$. Instead, it is most likely due to Inverse Compton scattering, which is commonly seen toward HzRGs \citep[e.g.][]{scharf03,overzier05}. For a typical massive cluster of $M\sim10^{15}$ $M_\odot$, the most massive progenitor that is part of its merger tree has a mass of about $M\sim10^{13}$ $M_\odot$ at $z\sim4$ \citep{chiang13}, which we will assume is a reasonable estimate for the halo mass of the TNJ1338 radio galaxy given what we know about its stellar mass. 

An important problem in the study of cluster formation is the origin of the entropy floor of the intracluster medium \citep{tozzi01}. \citet{roy24} showed that, despite the high mass outflow rates observed, the radio jet kinetic energy couples only weakly to the surrounding ISM. In the context of the massive (forming) dark matter halo, it is possible that the remaining energy contributes to heating of the surrounding medium as required by ICM preheating models, provided that the jet couples to the halo gas efficiently.  

Several other protoclusters have been found at $z\approx4$ \citep[e.g.][]{dey16,miller18,lemaux18}. 
Recently, \citet{chapman24} detected a radio AGN as part of the central core of galaxies in the protocluster SPT2349--56 at $z=4.3$. Although the precise host of the radio emission has not yet been identified within a region of 2\arcsec$\times$2\arcsec\ that contains 3 merging galaxies with very high stellar or dynamical masses, SPT2349--56 is therefore another protocluster with a radio-loud AGN at its center. Although the radio luminosity of SPT2349--56 is $\sim500\times$\ lower than that of TNJ1338, \citet{chapman24} estimate that the high power contained within the radio jets would be  sufficient to offset the cooling of hot gas within its inferred virial radius. 

Like TNJ1338, SPT2349--56 also has a luminous extended halo of \lya\ emission, indicating that luminous radio AGN may be responsible for distributing the gas to large scales and providing a source of photoionization. Another interesting and related aspect involves the origin of the metals in the extended ICM. Simulations have shown that a significant fraction of the Fe-rich gas in the ICM of local clusters originates from gas that was enriched at $z>2$ and brought in through accretion and merging between progenitor halos \citep{biffi18}. The significant mass outflow rates estimated by \citet{roy24} could be a source of such metal-enriched gas. 

Studying the centers of protoclusters like TNJ1338 and SPT2349--56 thus offer important insights not only on AGN feedback in massive early galaxies, but also on their connection to the development of the larger-scale cosmic web around them.

\section{Summary}
\label{sec:summary}

In this paper we present \emph{JWST} NIRSpec integral-field spectroscopic observations of TNJ1338$-$1942, a remarkably complex radio-loud AGN at $z=4.1$. The data cube has revealed both continuum and strong emission lines in the radio AGN, as well as spatially extended line emission along the radio jet axis tracing highly ionized conditions in the system. 

From the 2D emission line maps, we have identified five strong emission line regions in the field: TNJ1338-a, the host AGN, TNJ1338-b which is offset to the north west and presents extremely strong emission lines, TNJ1338-west which coincides with the apex of the previously identified \lya\ `wedge' structure, TNJ1338-east which is on the opposite side of the wedge and the radio axis, and TNJ1338-northwest which closely coincides with the peak of the radio emission.

A detailed analysis of the kinematics of the \ha+\nii\ complex and the \oiii\ doublet has revealed the presence of galactic scale outflows driven by the AGN, and highly disturbed gas distribution. The presence of broad lines of comparable widths in both permitted and forbidden lines at TNJ1338-a rules out an obvious broad-line region (BLR) arising from black hole accretion.

Using the classical BPT and VO87 line ratio diagnostics, we find that none of the strong line emitting regions are predominantly photoionized by star-formation activity, and must be ionized by the AGN with a possible contribution from fast radiative shocks. A 2D map of the \oiii/\oii\ (O32) ratio reveals a complex structure, with peaks at the locations of TNJ1338-a and b, indicating generally high ionization parameters and possibly multiple ionizing sources in the field. The (\oii+\oiii)/\hb\ (R23) map shows that the level of oxygen enrichment is relatively constant across the system, and we highlight a clear lack of metallicity gradients along the jet axis. Overall, we find that the field is highly ionized, with no obvious star-formation dominated regions seen in the analysis. 

Extremely high ionization potential \nev\ emission is seen prominently at the location of TNJ1338-a, as may be expected from bright AGN-driven photoionization. Interestingly, point-like \nev\ emission, albeit with relatively low significance ($\sim2\sigma$) is also seen at the location of TNJ1338-b. Using the \nev/\neiii\ line ratio along with predictions from various shock-driven photoionization models, we find that the Ne line ratios of both components are consistent with either photoionization by fast ($v>600$\,\kms) shocks, or by AGN. However, the line ratios do seem to rule out photoionization due to slower shocks in this region. 

Further, the \oiii\ intensity and kinematic maps produced by decomposing the line into broad and narrow component on a spaxel-by-spaxel basis have revealed prominent narrow line emission from all five emission line regions, with broad line emission appearing to align along the radio jet axis. This potentially traces the turbulence induced in the gas by the AGN jet as it traverses the medium. The velocity map reveals that the broadest line widths are seen west and north-west of the system close to the radio hotspot. These locations also present the highest outflow velocities.

These observations have helped shed light on the true nature of TNJ1338$-$1942, confirming that TNJ1338-a is likely the bright, obscured type-2 AGN photoionizing its surroundings and driving the radio jets observed. However, a combination of our findings have also led to the possibility that TNJ1338$-$1942 may host a second AGN, at the location of TNJ1338-b. The secondary AGN could trace another galaxy with an active SMBH that is merging with the main system as seen in other dual AGN systems \citep{comerford09,comerford14} or it could be due to the presence of a gravitational wave recoiling BH \citep{chiaberge17,chiaberge18}. The latter could offer an attractive scenario especially given the evidence for significant diffuse stellar mass but not a clear host galaxy around what would be the secondary AGN. In both the dual AGN and recoiling BH case, however, these findings would trace the process of merging of galaxies and BH growth.

The dual AGN scenario also explains a longstanding puzzle related to the origin of a large conical structure bright in \lya\ perpendicular to the radio axis. If this interpretation is correct, TNJ1338 would be an exceptional dual AGN system known at $z>4$. Recent \emph{JWST} observations, also with NIRSpec, have suggested that a substantial fraction of high redshift galaxies may contain dual AGN systems \citep{perna23a,perna23b}. These results are based on the same BPT line ratio diagnostics as performed in Section \ref{sec:ionization}. Although none of those sources is known to be radio loud, we note that we would have easily reached a similar conclusion had we not considered the complications due to the presence of the radio jet in TNJ1338. 

Our observations highlight the revolutionary capabilities of the NIRSpec IFU, and have led to new insights into the physical and chemical nature of TNJ1338$-$1942, revealing a highly complex and interesting radio AGN system in the early Universe. Detailed studies of such systems are essential to gain a better understanding of the assembly of some of the most massive and energetic systems across redshifts.

\section*{Acknowledgements}

AS would like to thank Chiara Circosta, Harley Katz and Richard Ellis for useful discussions. AS and AJC acknowledge funding from the ``FirstGalaxies'' Advanced Grant from the European Research Council (ERC) under the European Union's Horizon 2020 research and innovation programme (Grant agreement No. 789056). RO was supported by a productivity grant (302981/2019-5) from the National Council for Scientific and Technological Development (CNPq). MVM acknowledges support by grant Nr. PID2021-124665NB-I00  by the Spanish Ministry of Science and Innovation/State Agency of Research MCIN/AEI/ 10.13039/501100011033 and by ``ERDF A way of making Europe''. KJD acknowledges support from the STFC through an Ernest Rutherford Fellowship (grant number ST/W003120/1). SEIB is supported by the Deutsche Forschungsgemeinschaft (DFG) through Emmy Noether grant number BO 5771/1-1. KEG's research has received funding from the HUN-REN Hungarian Research Network. This research was funded by the Hungarian National Research, Development and Innovation Office (NKFIH), grant number OTKA K134213.

This work is based on observations made with the NASA/ESA/CSA James Webb Space Telescope. The data were obtained from the Mikulski Archive for Space Telescopes (MAST) at the Space Telescope Science Institute, which is operated by the Association of Universities for Research in Astronomy, Inc., under NASA contract NAS 5-03127 for JWST. These observations are associated with program 1964.


\section*{Data Availability}

The raw data from this observing program will become available on MAST 12 months after the date of observation, once the proprietary period runs out. In the mean time, access to derived data products will be provided by the authors upon reasonable written request.



\bibliographystyle{mnras}
\bibliography{scibib} 

\begin{thebibliography}{}
\makeatletter
\relax
\def\mn@urlcharsother{\let\do\@makeother \do\$\do\&\do\#\do\^\do\_\do\%\do\~}
\def\mn@doi{\begingroup\mn@urlcharsother \@ifnextchar [ {\mn@doi@}
  {\mn@doi@[]}}
\def\mn@doi@[#1]#2{\def\@tempa{#1}\ifx\@tempa\@empty \href
  {http://dx.doi.org/#2} {doi:#2}\else \href {http://dx.doi.org/#2} {#1}\fi
  \endgroup}
\def\mn@eprint#1#2{\mn@eprint@#1:#2::\@nil}
\def\mn@eprint@arXiv#1{\href {http://arxiv.org/abs/#1} {{\tt arXiv:#1}}}
\def\mn@eprint@dblp#1{\href {http://dblp.uni-trier.de/rec/bibtex/#1.xml}
  {dblp:#1}}
\def\mn@eprint@#1:#2:#3:#4\@nil{\def\@tempa {#1}\def\@tempb {#2}\def\@tempc
  {#3}\ifx \@tempc \@empty \let \@tempc \@tempb \let \@tempb \@tempa \fi \ifx
  \@tempb \@empty \def\@tempb {arXiv}\fi \@ifundefined
  {mn@eprint@\@tempb}{\@tempb:\@tempc}{\expandafter \expandafter \csname
  mn@eprint@\@tempb\endcsname \expandafter{\@tempc}}}

\bibitem[\protect\citeauthoryear{{Allen}, {Groves}, {Dopita}, {Sutherland}  \&
  {Kewley}}{{Allen} et~al.}{2008}]{allen08}
{Allen} M.~G.,  {Groves} B.~A.,  {Dopita} M.~A.,  {Sutherland} R.~S.,
  {Kewley} L.~J.,  2008, \mn@doi [\apjs] {10.1086/589652}, \href
  {https://ui.adsabs.harvard.edu/abs/2008ApJS..178...20A} {178, 20}

\bibitem[\protect\citeauthoryear{{Baldwin}, {Phillips}  \&
  {Terlevich}}{{Baldwin} et~al.}{1981}]{baldwin81}
{Baldwin} J.~A.,  {Phillips} M.~M.,   {Terlevich} R.,  1981, \mn@doi [\pasp]
  {10.1086/130766}, \href
  {https://ui.adsabs.harvard.edu/abs/1981PASP...93....5B} {93, 5}

\bibitem[\protect\citeauthoryear{{B{\"a}r}, {Weigel}, {Sartori}, {Oh}, {Koss}
  \& {Schawinski}}{{B{\"a}r} et~al.}{2017}]{baer2017}
{B{\"a}r} R.~E.,  {Weigel} A.~K.,  {Sartori} L.~F.,  {Oh} K.,  {Koss} M.,
  {Schawinski} K.,  2017, \mn@doi [\mnras] {10.1093/mnras/stw3283}, \href
  {https://ui.adsabs.harvard.edu/abs/2017MNRAS.466.2879B} {466, 2879}

\bibitem[\protect\citeauthoryear{{Begelman} \& {Cioffi}}{{Begelman} \&
  {Cioffi}}{1989}]{begelman89}
{Begelman} M.~C.,  {Cioffi} D.~F.,  1989, \mn@doi [\apjl] {10.1086/185542},
  \href {https://ui.adsabs.harvard.edu/abs/1989ApJ...345L..21B} {345, L21}

\bibitem[\protect\citeauthoryear{{Best}, {R{\"o}ttgering}  \& {Longair}}{{Best}
  et~al.}{2000}]{best00}
{Best} P.~N.,  {R{\"o}ttgering} H.~J.~A.,   {Longair} M.~S.,  2000, \mn@doi
  [\mnras] {10.1046/j.1365-8711.2000.03028.x}, \href
  {https://ui.adsabs.harvard.edu/abs/2000MNRAS.311...23B} {311, 23}

\bibitem[\protect\citeauthoryear{{Biffi}, {Planelles}, {Borgani}, {Rasia},
  {Murante}, {Fabjan}  \& {Gaspari}}{{Biffi} et~al.}{2018}]{biffi18}
{Biffi} V.,  {Planelles} S.,  {Borgani} S.,  {Rasia} E.,  {Murante} G.,
  {Fabjan} D.,   {Gaspari} M.,  2018, \mn@doi [\mnras] {10.1093/mnras/sty363},
  \href {https://ui.adsabs.harvard.edu/abs/2018MNRAS.476.2689B} {476, 2689}

\bibitem[\protect\citeauthoryear{{Blandford} \& {Znajek}}{{Blandford} \&
  {Znajek}}{1977}]{blandford77}
{Blandford} R.~D.,  {Znajek} R.~L.,  1977, \mn@doi [\mnras]
  {10.1093/mnras/179.3.433}, \href
  {https://ui.adsabs.harvard.edu/abs/1977MNRAS.179..433B} {179, 433}

\bibitem[\protect\citeauthoryear{{Carniani} et~al.,}{{Carniani}
  et~al.}{2023}]{carniani23}
{Carniani} S.,  et~al., 2023, \mn@doi [arXiv e-prints]
  {10.48550/arXiv.2306.11801}, \href
  {https://ui.adsabs.harvard.edu/abs/2023arXiv230611801C} {p. arXiv:2306.11801}

\bibitem[\protect\citeauthoryear{{Chambers}, {Miley}  \& {van
  Breugel}}{{Chambers} et~al.}{1990}]{chambers90}
{Chambers} K.~C.,  {Miley} G.~K.,   {van Breugel} W.~J.~M.,  1990, \mn@doi
  [\apj] {10.1086/169316}, \href
  {https://ui.adsabs.harvard.edu/abs/1990ApJ...363...21C} {363, 21}

\bibitem[\protect\citeauthoryear{{Chapman} et~al.,}{{Chapman}
  et~al.}{2024}]{chapman24}
{Chapman} S.~C.,  et~al., 2024, \mn@doi [\apj] {10.3847/1538-4357/ad0b77},
  \href {https://ui.adsabs.harvard.edu/abs/2024ApJ...961..120C} {961, 120}

\bibitem[\protect\citeauthoryear{{Chiaberge} \& {Marconi}}{{Chiaberge} \&
  {Marconi}}{2011}]{chiaberge11}
{Chiaberge} M.,  {Marconi} A.,  2011, \mn@doi [\mnras]
  {10.1111/j.1365-2966.2011.19079.x}, \href
  {https://ui.adsabs.harvard.edu/abs/2011MNRAS.416..917C} {416, 917}

\bibitem[\protect\citeauthoryear{{Chiaberge} et~al.,}{{Chiaberge}
  et~al.}{2017}]{chiaberge17}
{Chiaberge} M.,  et~al., 2017, \mn@doi [\aap] {10.1051/0004-6361/201629522},
  \href {https://ui.adsabs.harvard.edu/abs/2017A&A...600A..57C} {600, A57}

\bibitem[\protect\citeauthoryear{{Chiaberge}, {Tremblay}, {Capetti}  \&
  {Norman}}{{Chiaberge} et~al.}{2018}]{chiaberge18}
{Chiaberge} M.,  {Tremblay} G.~R.,  {Capetti} A.,   {Norman} C.,  2018, \mn@doi
  [\apj] {10.3847/1538-4357/aac48b}, \href
  {https://ui.adsabs.harvard.edu/abs/2018ApJ...861...56C} {861, 56}

\bibitem[\protect\citeauthoryear{{Chiang}, {Overzier}  \& {Gebhardt}}{{Chiang}
  et~al.}{2013}]{chiang13}
{Chiang} Y.-K.,  {Overzier} R.,   {Gebhardt} K.,  2013, \mn@doi [\apj]
  {10.1088/0004-637X/779/2/127}, \href
  {https://ui.adsabs.harvard.edu/abs/2013ApJ...779..127C} {779, 127}

\bibitem[\protect\citeauthoryear{{Clark}, {Axon}, {Tadhunter}, {Robinson}  \&
  {O'Brien}}{{Clark} et~al.}{1998}]{clark1998}
{Clark} N.~E.,  {Axon} D.~J.,  {Tadhunter} C.~N.,  {Robinson} A.,   {O'Brien}
  P.,  1998, \mn@doi [\apj] {10.1086/305225}, \href
  {https://ui.adsabs.harvard.edu/abs/1998ApJ...494..546C} {494, 546}

\bibitem[\protect\citeauthoryear{{Cleri} et~al.,}{{Cleri}
  et~al.}{2023}]{cleri23}
{Cleri} N.~J.,  et~al., 2023, \mn@doi [\apj] {10.3847/1538-4357/acde55}, \href
  {https://ui.adsabs.harvard.edu/abs/2023ApJ...953...10C} {953, 10}

\bibitem[\protect\citeauthoryear{{Comerford} \& {Greene}}{{Comerford} \&
  {Greene}}{2014}]{comerford14}
{Comerford} J.~M.,  {Greene} J.~E.,  2014, \mn@doi [\apj]
  {10.1088/0004-637X/789/2/112}, \href
  {https://ui.adsabs.harvard.edu/abs/2014ApJ...789..112C} {789, 112}

\bibitem[\protect\citeauthoryear{{Comerford}, {Griffith}, {Gerke}, {Cooper},
  {Newman}, {Davis}  \& {Stern}}{{Comerford} et~al.}{2009}]{comerford09}
{Comerford} J.~M.,  {Griffith} R.~L.,  {Gerke} B.~F.,  {Cooper} M.~C.,
  {Newman} J.~A.,  {Davis} M.,   {Stern} D.,  2009, \mn@doi [\apjl]
  {10.1088/0004-637X/702/1/L82}, \href
  {https://ui.adsabs.harvard.edu/abs/2009ApJ...702L..82C} {702, L82}

\bibitem[\protect\citeauthoryear{{De Breuck}, {van Breugel}, {Minniti},
  {Miley}, {R{\"o}ttgering}, {Stanford}  \& {Carilli}}{{De Breuck}
  et~al.}{1999}]{debreuck99}
{De Breuck} C.,  {van Breugel} W.,  {Minniti} D.,  {Miley} G.,
  {R{\"o}ttgering} H.,  {Stanford} S.~A.,   {Carilli} C.,  1999, \mn@doi [\aap]
  {10.48550/arXiv.astro-ph/9909178}, \href
  {https://ui.adsabs.harvard.edu/abs/1999A&A...352L..51D} {352, L51}

\bibitem[\protect\citeauthoryear{{De Breuck}, {van Breugel}, {R{\"o}ttgering}
  \& {Miley}}{{De Breuck} et~al.}{2000a}]{debreuck00a}
{De Breuck} C.,  {van Breugel} W.,  {R{\"o}ttgering} H.~J.~A.,   {Miley} G.,
  2000a, \mn@doi [\aaps] {10.1051/aas:2000181}, \href
  {https://ui.adsabs.harvard.edu/abs/2000A&AS..143..303D} {143, 303}

\bibitem[\protect\citeauthoryear{{De Breuck}, {R{\"o}ttgering}, {Miley}, {van
  Breugel}  \& {Best}}{{De Breuck} et~al.}{2000b}]{debreuck00b}
{De Breuck} C.,  {R{\"o}ttgering} H.,  {Miley} G.,  {van Breugel} W.,   {Best}
  P.,  2000b, \mn@doi [\aap] {10.48550/arXiv.astro-ph/0008264}, \href
  {https://ui.adsabs.harvard.edu/abs/2000A&A...362..519D} {362, 519}

\bibitem[\protect\citeauthoryear{{De Breuck} et~al.,}{{De Breuck}
  et~al.}{2004}]{debreuck04}
{De Breuck} C.,  et~al., 2004, \mn@doi [\aap] {10.1051/0004-6361:20035885},
  \href {https://ui.adsabs.harvard.edu/abs/2004A&A...424....1D} {424, 1}

\bibitem[\protect\citeauthoryear{{De Breuck} et~al.,}{{De Breuck}
  et~al.}{2010}]{debreuck10}
{De Breuck} C.,  et~al., 2010, \mn@doi [\apj] {10.1088/0004-637X/725/1/36},
  \href {https://ui.adsabs.harvard.edu/abs/2010ApJ...725...36D} {725, 36}

\bibitem[\protect\citeauthoryear{{Dey}, {van Breugel}, {Vacca}  \&
  {Antonucci}}{{Dey} et~al.}{1997}]{dey97}
{Dey} A.,  {van Breugel} W.,  {Vacca} W.~D.,   {Antonucci} R.,  1997, \mn@doi
  [\apj] {10.1086/304911}, \href
  {https://ui.adsabs.harvard.edu/abs/1997ApJ...490..698D} {490, 698}

\bibitem[\protect\citeauthoryear{{Dey}, {Lee}, {Reddy}, {Cooper}, {Inami},
  {Hong}, {Gonzalez}  \& {Jannuzi}}{{Dey} et~al.}{2016}]{dey16}
{Dey} A.,  {Lee} K.-S.,  {Reddy} N.,  {Cooper} M.,  {Inami} H.,  {Hong} S.,
  {Gonzalez} A.~H.,   {Jannuzi} B.~T.,  2016, \mn@doi [\apj]
  {10.3847/0004-637X/823/1/11}, \href
  {https://ui.adsabs.harvard.edu/abs/2016ApJ...823...11D} {823, 11}

\bibitem[\protect\citeauthoryear{{Dopita} \& {Sutherland}}{{Dopita} \&
  {Sutherland}}{1996}]{dopita96}
{Dopita} M.~A.,  {Sutherland} R.~S.,  1996, \mn@doi [\apjs] {10.1086/192255},
  \href {https://ui.adsabs.harvard.edu/abs/1996ApJS..102..161D} {102, 161}

\bibitem[\protect\citeauthoryear{{Dors}}{{Dors}}{2021}]{dors2021}
{Dors} O.~L.,  2021, \mn@doi [\mnras] {10.1093/mnras/stab2166}, \href
  {https://ui.adsabs.harvard.edu/abs/2021MNRAS.507..466D} {507, 466}

\bibitem[\protect\citeauthoryear{{Drouart}, {Rocca-Volmerange}, {De Breuck},
  {Fioc}, {Lehnert}, {Seymour}, {Stern}  \& {Vernet}}{{Drouart}
  et~al.}{2016}]{drouart16}
{Drouart} G.,  {Rocca-Volmerange} B.,  {De Breuck} C.,  {Fioc} M.,  {Lehnert}
  M.,  {Seymour} N.,  {Stern} D.,   {Vernet} J.,  2016, \mn@doi [\aap]
  {10.1051/0004-6361/201526880}, \href
  {https://ui.adsabs.harvard.edu/abs/2016A&A...593A.109D} {593, A109}

\bibitem[\protect\citeauthoryear{{Duncan} et~al.,}{{Duncan}
  et~al.}{2023}]{duncan23}
{Duncan} K.~J.,  et~al., 2023, \mn@doi [\mnras] {10.1093/mnras/stad1267}, \href
  {https://ui.adsabs.harvard.edu/abs/2023MNRAS.522.4548D} {522, 4548}

\bibitem[\protect\citeauthoryear{{Emonts} et~al.,}{{Emonts}
  et~al.}{2016}]{emonts16}
{Emonts} B.~H.~C.,  et~al., 2016, \mn@doi [Science] {10.1126/science.aag0512},
  \href {https://ui.adsabs.harvard.edu/abs/2016Sci...354.1128E} {354, 1128}

\bibitem[\protect\citeauthoryear{{Enoki}, {Ishiyama}, {Kobayashi}  \&
  {Nagashima}}{{Enoki} et~al.}{2014}]{enoki14}
{Enoki} M.,  {Ishiyama} T.,  {Kobayashi} M. A.~R.,   {Nagashima} M.,  2014,
  \mn@doi [\apj] {10.1088/0004-637X/794/1/69}, \href
  {https://ui.adsabs.harvard.edu/abs/2014ApJ...794...69E} {794, 69}

\bibitem[\protect\citeauthoryear{{Fabian}}{{Fabian}}{2012}]{fabian12}
{Fabian} A.~C.,  2012, \mn@doi [\araa] {10.1146/annurev-astro-081811-125521},
  \href {https://ui.adsabs.harvard.edu/abs/2012ARA&A..50..455F} {50, 455}

\bibitem[\protect\citeauthoryear{{Falkendal} et~al.,}{{Falkendal}
  et~al.}{2019}]{falkendal19}
{Falkendal} T.,  et~al., 2019, \mn@doi [\aap] {10.1051/0004-6361/201732485},
  \href {https://ui.adsabs.harvard.edu/abs/2019A&A...621A..27F} {621, A27}

\bibitem[\protect\citeauthoryear{{Fan}, {Ba{\~n}ados}  \& {Simcoe}}{{Fan}
  et~al.}{2023}]{fan23}
{Fan} X.,  {Ba{\~n}ados} E.,   {Simcoe} R.~A.,  2023, \mn@doi [\araa]
  {10.1146/annurev-astro-052920-102455}, \href
  {https://ui.adsabs.harvard.edu/abs/2023ARA&A..61..373F} {61, 373}

\bibitem[\protect\citeauthoryear{{Gilli}, {Vignali}, {Mignoli}, {Iwasawa},
  {Comastri}  \& {Zamorani}}{{Gilli} et~al.}{2010}]{gilli10}
{Gilli} R.,  {Vignali} C.,  {Mignoli} M.,  {Iwasawa} K.,  {Comastri} A.,
  {Zamorani} G.,  2010, \mn@doi [\aap] {10.1051/0004-6361/201014039}, \href
  {https://ui.adsabs.harvard.edu/abs/2010A&A...519A..92G} {519, A92}

\bibitem[\protect\citeauthoryear{{Groves}, {Heckman}  \& {Kauffmann}}{{Groves}
  et~al.}{2006}]{groves06}
{Groves} B.~A.,  {Heckman} T.~M.,   {Kauffmann} G.,  2006, \mn@doi [\mnras]
  {10.1111/j.1365-2966.2006.10812.x}, \href
  {https://ui.adsabs.harvard.edu/abs/2006MNRAS.371.1559G} {371, 1559}

\bibitem[\protect\citeauthoryear{{Hardcastle} \& {Croston}}{{Hardcastle} \&
  {Croston}}{2020}]{hardcastle20}
{Hardcastle} M.~J.,  {Croston} J.~H.,  2020, \mn@doi [\nar]
  {10.1016/j.newar.2020.101539}, \href
  {https://ui.adsabs.harvard.edu/abs/2020NewAR..8801539H} {88, 101539}

\bibitem[\protect\citeauthoryear{{Harikane} et~al.,}{{Harikane}
  et~al.}{2023}]{harikane23}
{Harikane} Y.,  et~al., 2023, \mn@doi [\apj] {10.3847/1538-4357/ad029e}, \href
  {https://ui.adsabs.harvard.edu/abs/2023ApJ...959...39H} {959, 39}

\bibitem[\protect\citeauthoryear{{Hatch}, {Overzier}, {Kurk}, {Miley},
  {R{\"o}ttgering}  \& {Zirm}}{{Hatch} et~al.}{2009}]{hatch09}
{Hatch} N.~A.,  {Overzier} R.~A.,  {Kurk} J.~D.,  {Miley} G.~K.,
  {R{\"o}ttgering} H.~J.~A.,   {Zirm} A.~W.,  2009, \mn@doi [\mnras]
  {10.1111/j.1365-2966.2009.14525.x}, \href
  {https://ui.adsabs.harvard.edu/abs/2009MNRAS.395..114H} {395, 114}

\bibitem[\protect\citeauthoryear{{Heckman} \& {Best}}{{Heckman} \&
  {Best}}{2014}]{heckman14}
{Heckman} T.~M.,  {Best} P.~N.,  2014, \mn@doi [\araa]
  {10.1146/annurev-astro-081913-035722}, \href
  {https://ui.adsabs.harvard.edu/abs/2014ARA&A..52..589H} {52, 589}

\bibitem[\protect\citeauthoryear{{Heckman}, {Armus}  \& {Miley}}{{Heckman}
  et~al.}{1990}]{heckman90}
{Heckman} T.~M.,  {Armus} L.,   {Miley} G.~K.,  1990, \mn@doi [\apjs]
  {10.1086/191522}, \href
  {https://ui.adsabs.harvard.edu/abs/1990ApJS...74..833H} {74, 833}

\bibitem[\protect\citeauthoryear{{Heckman}, {Lehnert}, {Miley}  \& {van
  Breugel}}{{Heckman} et~al.}{1991}]{heckman91}
{Heckman} T.~M.,  {Lehnert} M.~D.,  {Miley} G.~K.,   {van Breugel} W.,  1991,
  \mn@doi [\apj] {10.1086/170660}, \href
  {https://ui.adsabs.harvard.edu/abs/1991ApJ...381..373H} {381, 373}

\bibitem[\protect\citeauthoryear{{Heckman}, {Kauffmann}, {Brinchmann},
  {Charlot}, {Tremonti}  \& {White}}{{Heckman} et~al.}{2004}]{heckman04}
{Heckman} T.~M.,  {Kauffmann} G.,  {Brinchmann} J.,  {Charlot} S.,  {Tremonti}
  C.,   {White} S. D.~M.,  2004, \mn@doi [\apj] {10.1086/422872}, \href
  {https://ui.adsabs.harvard.edu/abs/2004ApJ...613..109H} {613, 109}

\bibitem[\protect\citeauthoryear{{Humphrey}, {Villar-Mart{\'\i}n}, {Vernet},
  {Fosbury}, {di Serego Alighieri}  \& {Binette}}{{Humphrey}
  et~al.}{2008}]{humphrey08}
{Humphrey} A.,  {Villar-Mart{\'\i}n} M.,  {Vernet} J.,  {Fosbury} R.,  {di
  Serego Alighieri} S.,   {Binette} L.,  2008, \mn@doi [\mnras]
  {10.1111/j.1365-2966.2007.12506.x}, \href
  {https://ui.adsabs.harvard.edu/abs/2008MNRAS.383...11H} {383, 11}

\bibitem[\protect\citeauthoryear{{Intema}, {Venemans}, {Kurk}, {Ouchi},
  {Kodama}, {R{\"o}ttgering}, {Miley}  \& {Overzier}}{{Intema}
  et~al.}{2006}]{intema06}
{Intema} H.~T.,  {Venemans} B.~P.,  {Kurk} J.~D.,  {Ouchi} M.,  {Kodama} T.,
  {R{\"o}ttgering} H.~J.~A.,  {Miley} G.~K.,   {Overzier} R.~A.,  2006, \mn@doi
  [\aap] {10.1051/0004-6361:20064812}, \href
  {https://ui.adsabs.harvard.edu/abs/2006A&A...456..433I} {456, 433}

\bibitem[\protect\citeauthoryear{{Izotov}, {Thuan}  \& {Privon}}{{Izotov}
  et~al.}{2012}]{izotov12}
{Izotov} Y.~I.,  {Thuan} T.~X.,   {Privon} G.,  2012, \mn@doi [\mnras]
  {10.1111/j.1365-2966.2012.22051.x}, \href
  {https://ui.adsabs.harvard.edu/abs/2012MNRAS.427.1229I} {427, 1229}

\bibitem[\protect\citeauthoryear{{Jia}, {Ptak}, {Heckman}, {Overzier},
  {Hornschemeier}  \& {LaMassa}}{{Jia} et~al.}{2011}]{jia11}
{Jia} J.,  {Ptak} A.,  {Heckman} T.~M.,  {Overzier} R.~A.,  {Hornschemeier} A.,
    {LaMassa} S.~M.,  2011, \mn@doi [\apj] {10.1088/0004-637X/731/1/55}, \href
  {https://ui.adsabs.harvard.edu/abs/2011ApJ...731...55J} {731, 55}

\bibitem[\protect\citeauthoryear{{Kakkad} et~al.,}{{Kakkad}
  et~al.}{2020}]{kakkad20}
{Kakkad} D.,  et~al., 2020, \mn@doi [\aap] {10.1051/0004-6361/202038551}, \href
  {https://ui.adsabs.harvard.edu/abs/2020A&A...642A.147K} {642, A147}

\bibitem[\protect\citeauthoryear{{Kewley}, {Groves}, {Kauffmann}  \&
  {Heckman}}{{Kewley} et~al.}{2006}]{kewley06}
{Kewley} L.~J.,  {Groves} B.,  {Kauffmann} G.,   {Heckman} T.,  2006, \mn@doi
  [\mnras] {10.1111/j.1365-2966.2006.10859.x}, \href
  {https://ui.adsabs.harvard.edu/abs/2006MNRAS.372..961K} {372, 961}

\bibitem[\protect\citeauthoryear{{Kewley}, {Dopita}, {Leitherer}, {Dav{\'e}},
  {Yuan}, {Allen}, {Groves}  \& {Sutherland}}{{Kewley}
  et~al.}{2013}]{kewley2013}
{Kewley} L.~J.,  {Dopita} M.~A.,  {Leitherer} C.,  {Dav{\'e}} R.,  {Yuan} T.,
  {Allen} M.,  {Groves} B.,   {Sutherland} R.,  2013, \mn@doi [\apj]
  {10.1088/0004-637X/774/2/100}, \href
  {https://ui.adsabs.harvard.edu/abs/2013ApJ...774..100K} {774, 100}

\bibitem[\protect\citeauthoryear{{Kolwa} et~al.,}{{Kolwa}
  et~al.}{2019}]{kolwa19}
{Kolwa} S.,  et~al., 2019, \mn@doi [\aap] {10.1051/0004-6361/201935437}, \href
  {https://ui.adsabs.harvard.edu/abs/2019A&A...625A.102K} {625, A102}

\bibitem[\protect\citeauthoryear{{Kormendy} \& {Ho}}{{Kormendy} \&
  {Ho}}{2013}]{kormendy13}
{Kormendy} J.,  {Ho} L.~C.,  2013, \mn@doi [\araa]
  {10.1146/annurev-astro-082708-101811}, \href
  {https://ui.adsabs.harvard.edu/abs/2013ARA&A..51..511K} {51, 511}

\bibitem[\protect\citeauthoryear{{Lemaux} et~al.,}{{Lemaux}
  et~al.}{2018}]{lemaux18}
{Lemaux} B.~C.,  et~al., 2018, \mn@doi [\aap] {10.1051/0004-6361/201730870},
  \href {https://ui.adsabs.harvard.edu/abs/2018A&A...615A..77L} {615, A77}

\bibitem[\protect\citeauthoryear{{Leung}, {Coil}, {Rupke}  \&
  {Perrotta}}{{Leung} et~al.}{2021}]{leung21}
{Leung} G. C.~K.,  {Coil} A.~L.,  {Rupke} D. S.~N.,   {Perrotta} S.,  2021,
  \mn@doi [\apj] {10.3847/1538-4357/abf4da}, \href
  {https://ui.adsabs.harvard.edu/abs/2021ApJ...914...17L} {914, 17}

\bibitem[\protect\citeauthoryear{{Magliocchetti}}{{Magliocchetti}}{2022}]{magliocchetti22}
{Magliocchetti} M.,  2022, \mn@doi [\aapr] {10.1007/s00159-022-00142-1}, \href
  {https://ui.adsabs.harvard.edu/abs/2022A&ARv..30....6M} {30, 6}

\bibitem[\protect\citeauthoryear{{Maiolino} et~al.,}{{Maiolino}
  et~al.}{2023}]{maiolino23}
{Maiolino} R.,  et~al., 2023, \mn@doi [arXiv e-prints]
  {10.48550/arXiv.2308.01230}, \href
  {https://ui.adsabs.harvard.edu/abs/2023arXiv230801230M} {p. arXiv:2308.01230}

\bibitem[\protect\citeauthoryear{{Man}, {Lehnert}, {Vernet}, {De Breuck}  \&
  {Falkendal}}{{Man} et~al.}{2019}]{man19}
{Man} A. W.~S.,  {Lehnert} M.~D.,  {Vernet} J. D.~R.,  {De Breuck} C.,
  {Falkendal} T.,  2019, \mn@doi [\aap] {10.1051/0004-6361/201834542}, \href
  {https://ui.adsabs.harvard.edu/abs/2019A&A...624A..81M} {624, A81}

\bibitem[\protect\citeauthoryear{{Marshall} et~al.,}{{Marshall}
  et~al.}{2023}]{mar23}
{Marshall} M.~A.,  et~al., 2023, \mn@doi [arXiv e-prints]
  {10.48550/arXiv.2302.04795}, \href
  {https://ui.adsabs.harvard.edu/abs/2023arXiv230204795M} {p. arXiv:2302.04795}

\bibitem[\protect\citeauthoryear{{McCarthy}}{{McCarthy}}{1993}]{mccarthy93}
{McCarthy} P.~J.,  1993, \mn@doi [\araa] {10.1146/annurev.aa.31.090193.003231},
  \href {https://ui.adsabs.harvard.edu/abs/1993ARA&A..31..639M} {31, 639}

\bibitem[\protect\citeauthoryear{{McCarthy}, {Spinrad}, {Djorgovski},
  {Strauss}, {van Breugel}  \& {Liebert}}{{McCarthy} et~al.}{1987}]{mccarthy87}
{McCarthy} P.~J.,  {Spinrad} H.,  {Djorgovski} S.,  {Strauss} M.~A.,  {van
  Breugel} W.,   {Liebert} J.,  1987, \mn@doi [\apjl] {10.1086/184951}, \href
  {https://ui.adsabs.harvard.edu/abs/1987ApJ...319L..39M} {319, L39}

\bibitem[\protect\citeauthoryear{{McConnell}, {Ma}, {Gebhardt}, {Wright},
  {Murphy}, {Lauer}, {Graham}  \& {Richstone}}{{McConnell}
  et~al.}{2011}]{mcconnell11}
{McConnell} N.~J.,  {Ma} C.-P.,  {Gebhardt} K.,  {Wright} S.~A.,  {Murphy}
  J.~D.,  {Lauer} T.~R.,  {Graham} J.~R.,   {Richstone} D.~O.,  2011, \mn@doi
  [\nat] {10.1038/nature10636}, \href
  {https://ui.adsabs.harvard.edu/abs/2011Natur.480..215M} {480, 215}

\bibitem[\protect\citeauthoryear{{McLure} \& {Dunlop}}{{McLure} \&
  {Dunlop}}{2004}]{mclure04}
{McLure} R.~J.,  {Dunlop} J.~S.,  2004, \mn@doi [\mnras]
  {10.1111/j.1365-2966.2004.08034.x}, \href
  {https://ui.adsabs.harvard.edu/abs/2004MNRAS.352.1390M} {352, 1390}

\bibitem[\protect\citeauthoryear{{Merloni}}{{Merloni}}{2004}]{merloni04}
{Merloni} A.,  2004, \mn@doi [\mnras] {10.1111/j.1365-2966.2004.08147.x}, \href
  {https://ui.adsabs.harvard.edu/abs/2004MNRAS.353.1035M} {353, 1035}

\bibitem[\protect\citeauthoryear{{Mignoli} et~al.,}{{Mignoli}
  et~al.}{2013}]{mignoli13}
{Mignoli} M.,  et~al., 2013, \mn@doi [\aap] {10.1051/0004-6361/201220846},
  \href {https://ui.adsabs.harvard.edu/abs/2013A&A...556A..29M} {556, A29}

\bibitem[\protect\citeauthoryear{{Miley} \& {De Breuck}}{{Miley} \& {De
  Breuck}}{2008}]{miley08}
{Miley} G.,  {De Breuck} C.,  2008, \mn@doi [\aapr]
  {10.1007/s00159-007-0008-z}, \href
  {https://ui.adsabs.harvard.edu/abs/2008A&ARv..15...67M} {15, 67}

\bibitem[\protect\citeauthoryear{{Miley} et~al.,}{{Miley}
  et~al.}{2004}]{miley04}
{Miley} G.~K.,  et~al., 2004, \mn@doi [\nat] {10.1038/nature02125}, \href
  {https://ui.adsabs.harvard.edu/abs/2004Natur.427...47M} {427, 47}

\bibitem[\protect\citeauthoryear{{Miley} et~al.,}{{Miley}
  et~al.}{2006}]{miley06}
{Miley} G.~K.,  et~al., 2006, \mn@doi [\apjl] {10.1086/508534}, \href
  {https://ui.adsabs.harvard.edu/abs/2006ApJ...650L..29M} {650, L29}

\bibitem[\protect\citeauthoryear{{Miller} et~al.,}{{Miller}
  et~al.}{2018}]{miller18}
{Miller} T.~B.,  et~al., 2018, \mn@doi [\nat] {10.1038/s41586-018-0025-2},
  \href {https://ui.adsabs.harvard.edu/abs/2018Natur.556..469M} {556, 469}

\bibitem[\protect\citeauthoryear{{Morais} et~al.,}{{Morais}
  et~al.}{2017}]{morais17}
{Morais} S.~G.,  et~al., 2017, \mn@doi [\mnras] {10.1093/mnras/stw2926}, \href
  {https://ui.adsabs.harvard.edu/abs/2017MNRAS.465.2698M} {465, 2698}

\bibitem[\protect\citeauthoryear{{Neeleman} et~al.,}{{Neeleman}
  et~al.}{2021}]{neeleman21}
{Neeleman} M.,  et~al., 2021, \mn@doi [\apj] {10.3847/1538-4357/abe70f}, \href
  {https://ui.adsabs.harvard.edu/abs/2021ApJ...911..141N} {911, 141}

\bibitem[\protect\citeauthoryear{{Nesvadba}, {Lehnert}, {Eisenhauer},
  {Gilbert}, {Tecza}  \& {Abuter}}{{Nesvadba} et~al.}{2006}]{nesvadba06}
{Nesvadba} N.~P.~H.,  {Lehnert} M.~D.,  {Eisenhauer} F.,  {Gilbert} A.,
  {Tecza} M.,   {Abuter} R.,  2006, \mn@doi [\apj] {10.1086/507266}, \href
  {https://ui.adsabs.harvard.edu/abs/2006ApJ...650..693N} {650, 693}

\bibitem[\protect\citeauthoryear{{Nesvadba}, {De Breuck}, {Lehnert}, {Best},
  {Binette}  \& {Proga}}{{Nesvadba} et~al.}{2011}]{nesvadba11}
{Nesvadba} N.~P.~H.,  {De Breuck} C.,  {Lehnert} M.~D.,  {Best} P.~N.,
  {Binette} L.,   {Proga} D.,  2011, \mn@doi [\aap]
  {10.1051/0004-6361/201014960}, \href
  {https://ui.adsabs.harvard.edu/abs/2011A&A...525A..43N} {525, A43}

\bibitem[\protect\citeauthoryear{{Nesvadba}, {De Breuck}, {Lehnert}, {Best}  \&
  {Collet}}{{Nesvadba} et~al.}{2017a}]{nesvadba17a}
{Nesvadba} N.~P.~H.,  {De Breuck} C.,  {Lehnert} M.~D.,  {Best} P.~N.,
  {Collet} C.,  2017a, \mn@doi [\aap] {10.1051/0004-6361/201528040}, \href
  {https://ui.adsabs.harvard.edu/abs/2017A&A...599A.123N} {599, A123}

\bibitem[\protect\citeauthoryear{{Nesvadba}, {Drouart}, {De Breuck}, {Best},
  {Seymour}  \& {Vernet}}{{Nesvadba} et~al.}{2017b}]{nesvadba17b}
{Nesvadba} N.~P.~H.,  {Drouart} G.,  {De Breuck} C.,  {Best} P.,  {Seymour} N.,
    {Vernet} J.,  2017b, \mn@doi [\aap] {10.1051/0004-6361/201629357}, \href
  {https://ui.adsabs.harvard.edu/abs/2017A&A...600A.121N} {600, A121}

\bibitem[\protect\citeauthoryear{{Netzer}, {Lira}, {Trakhtenbrot}, {Shemmer}
  \& {Cury}}{{Netzer} et~al.}{2007}]{netzer07}
{Netzer} H.,  {Lira} P.,  {Trakhtenbrot} B.,  {Shemmer} O.,   {Cury} I.,  2007,
  \mn@doi [\apj] {10.1086/523035}, \href
  {https://ui.adsabs.harvard.edu/abs/2007ApJ...671.1256N} {671, 1256}

\bibitem[\protect\citeauthoryear{{Osterbrock} \& {Ferland}}{{Osterbrock} \&
  {Ferland}}{2006}]{ost06}
{Osterbrock} D.~E.,  {Ferland} G.~J.,  2006, {Astrophysics of gaseous nebulae
  and active galactic nuclei}

\bibitem[\protect\citeauthoryear{{Overzier}}{{Overzier}}{2016}]{overzier16}
{Overzier} R.~A.,  2016, \mn@doi [\aapr] {10.1007/s00159-016-0100-3}, \href
  {https://ui.adsabs.harvard.edu/abs/2016A&ARv..24...14O} {24, 14}

\bibitem[\protect\citeauthoryear{{Overzier}, {Harris}, {Carilli}, {Pentericci},
  {R{\"o}ttgering}  \& {Miley}}{{Overzier} et~al.}{2005}]{overzier05}
{Overzier} R.~A.,  {Harris} D.~E.,  {Carilli} C.~L.,  {Pentericci} L.,
  {R{\"o}ttgering} H.~J.~A.,   {Miley} G.~K.,  2005, \mn@doi [\aap]
  {10.1051/0004-6361:20041657}, \href
  {https://ui.adsabs.harvard.edu/abs/2005A&A...433...87O} {433, 87}

\bibitem[\protect\citeauthoryear{{Overzier} et~al.,}{{Overzier}
  et~al.}{2008}]{overzier08}
{Overzier} R.~A.,  et~al., 2008, \mn@doi [\apj] {10.1086/524342}, \href
  {https://ui.adsabs.harvard.edu/abs/2008ApJ...673..143O} {673, 143}

\bibitem[\protect\citeauthoryear{{Overzier} et~al.,}{{Overzier}
  et~al.}{2009}]{overzier09}
{Overzier} R.~A.,  et~al., 2009, \mn@doi [\apj] {10.1088/0004-637X/704/1/548},
  \href {https://ui.adsabs.harvard.edu/abs/2009ApJ...704..548O} {704, 548}

\bibitem[\protect\citeauthoryear{{Pentericci}, {R{\"o}ttgering}, {Miley},
  {Spinrad}, {McCarthy}, {van Breugel}  \& {Macchetto}}{{Pentericci}
  et~al.}{1998}]{pentericci98}
{Pentericci} L.,  {R{\"o}ttgering} H.~J.~A.,  {Miley} G.~K.,  {Spinrad} H.,
  {McCarthy} P.~J.,  {van Breugel} W.~J.~M.,   {Macchetto} F.,  1998, \mn@doi
  [\apj] {10.1086/306087}, \href
  {https://ui.adsabs.harvard.edu/abs/1998ApJ...504..139P} {504, 139}

\bibitem[\protect\citeauthoryear{{Pentericci}, {R{\"o}ttgering}, {Miley},
  {McCarthy}, {Spinrad}, {van Breugel}  \& {Macchetto}}{{Pentericci}
  et~al.}{1999}]{pentericci99}
{Pentericci} L.,  {R{\"o}ttgering} H.~J.~A.,  {Miley} G.~K.,  {McCarthy} P.,
  {Spinrad} H.,  {van Breugel} W.~J.~M.,   {Macchetto} F.,  1999, \mn@doi
  [\aap] {10.48550/arXiv.astro-ph/9809056}, \href
  {https://ui.adsabs.harvard.edu/abs/1999A&A...341..329P} {341, 329}

\bibitem[\protect\citeauthoryear{{Pentericci}, {Van Reeven}, {Carilli},
  {R{\"o}ttgering}  \& {Miley}}{{Pentericci} et~al.}{2000}]{pentericci00}
{Pentericci} L.,  {Van Reeven} W.,  {Carilli} C.~L.,  {R{\"o}ttgering}
  H.~J.~A.,   {Miley} G.~K.,  2000, \mn@doi [\aaps] {10.1051/aas:2000104},
  \href {https://ui.adsabs.harvard.edu/abs/2000A&AS..145..121P} {145, 121}

\bibitem[\protect\citeauthoryear{{Pentericci}, {McCarthy}, {R{\"o}ttgering},
  {Miley}, {van Breugel}  \& {Fosbury}}{{Pentericci}
  et~al.}{2001}]{pentericci01}
{Pentericci} L.,  {McCarthy} P.~J.,  {R{\"o}ttgering} H.~J.~A.,  {Miley} G.~K.,
   {van Breugel} W.~J.~M.,   {Fosbury} R.,  2001, \mn@doi [\apjs]
  {10.1086/321781}, \href
  {https://ui.adsabs.harvard.edu/abs/2001ApJS..135...63P} {135, 63}

\bibitem[\protect\citeauthoryear{{Perna} et~al.,}{{Perna}
  et~al.}{2023a}]{perna23b}
{Perna} M.,  et~al., 2023a, \mn@doi [arXiv e-prints]
  {10.48550/arXiv.2310.03067}, \href
  {https://ui.adsabs.harvard.edu/abs/2023arXiv231003067P} {p. arXiv:2310.03067}

\bibitem[\protect\citeauthoryear{{Perna} et~al.,}{{Perna}
  et~al.}{2023b}]{perna23a}
{Perna} M.,  et~al., 2023b, \mn@doi [\aap] {10.1051/0004-6361/202346649}, \href
  {https://ui.adsabs.harvard.edu/abs/2023A&A...679A..89P} {679, A89}

\bibitem[\protect\citeauthoryear{{Planck Collaboration} et~al.,}{{Planck
  Collaboration} et~al.}{2020}]{planck20}
{Planck Collaboration} et~al., 2020, \mn@doi [\aap]
  {10.1051/0004-6361/201833910}, \href
  {https://ui.adsabs.harvard.edu/abs/2020A&A...641A...6P} {641, A6}

\bibitem[\protect\citeauthoryear{{Poitevineau}, {Castignani}  \&
  {Combes}}{{Poitevineau} et~al.}{2023}]{poitevineau23}
{Poitevineau} R.,  {Castignani} G.,   {Combes} F.,  2023, \mn@doi [\aap]
  {10.1051/0004-6361/202244560}, \href
  {https://ui.adsabs.harvard.edu/abs/2023A&A...672A.164P} {672, A164}

\bibitem[\protect\citeauthoryear{{Rees}}{{Rees}}{1984}]{rees84}
{Rees} M.~J.,  1984, \mn@doi [\araa] {10.1146/annurev.aa.22.090184.002351},
  \href {https://ui.adsabs.harvard.edu/abs/1984ARA&A..22..471R} {22, 471}

\bibitem[\protect\citeauthoryear{{Retana-Montenegro} \&
  {R{\"o}ttgering}}{{Retana-Montenegro} \&
  {R{\"o}ttgering}}{2017}]{retana-montenegro17}
{Retana-Montenegro} E.,  {R{\"o}ttgering} H.~J.~A.,  2017, \mn@doi [\aap]
  {10.1051/0004-6361/201526433}, \href
  {https://ui.adsabs.harvard.edu/abs/2017A&A...600A..97R} {600, A97}

\bibitem[\protect\citeauthoryear{{Rigby}, {Argyle}, {Best}, {Rosario}  \&
  {R{\"o}ttgering}}{{Rigby} et~al.}{2015}]{rigby15}
{Rigby} E.~E.,  {Argyle} J.,  {Best} P.~N.,  {Rosario} D.,   {R{\"o}ttgering}
  H.~J.~A.,  2015, \mn@doi [\aap] {10.1051/0004-6361/201526475}, \href
  {https://ui.adsabs.harvard.edu/abs/2015A&A...581A..96R} {581, A96}

\bibitem[\protect\citeauthoryear{{Roy}, {Heckman}, {Overzier}, {Saxena}  \& {et
  al.}}{{Roy} et~al.}{2024}]{roy24}
{Roy} N.,  {Heckman} T.~H.,  {Overzier} R.~A.,  {Saxena} A.,   {et al.} 2024,
  submitted.

\bibitem[\protect\citeauthoryear{{Rupke}, {Veilleux}  \& {Sanders}}{{Rupke}
  et~al.}{2005}]{rupke05}
{Rupke} D.~S.,  {Veilleux} S.,   {Sanders} D.~B.,  2005, \mn@doi [\apj]
  {10.1086/444451}, \href
  {https://ui.adsabs.harvard.edu/abs/2005ApJ...632..751R} {632, 751}

\bibitem[\protect\citeauthoryear{{Sabater} et~al.,}{{Sabater}
  et~al.}{2019}]{sabater19}
{Sabater} J.,  et~al., 2019, \mn@doi [\aap] {10.1051/0004-6361/201833883},
  \href {https://ui.adsabs.harvard.edu/abs/2019A&A...622A..17S} {622, A17}

\bibitem[\protect\citeauthoryear{{Saito} et~al.,}{{Saito}
  et~al.}{2015}]{saito15}
{Saito} T.,  et~al., 2015, \mn@doi [\mnras] {10.1093/mnras/stu2538}, \href
  {https://ui.adsabs.harvard.edu/abs/2015MNRAS.447.3069S} {447, 3069}

\bibitem[\protect\citeauthoryear{{Santoro}, {Tadhunter}, {Baron}, {Morganti}
  \& {Holt}}{{Santoro} et~al.}{2020}]{santoro20}
{Santoro} F.,  {Tadhunter} C.,  {Baron} D.,  {Morganti} R.,   {Holt} J.,  2020,
  \mn@doi [\aap] {10.1051/0004-6361/202039077}, \href
  {https://ui.adsabs.harvard.edu/abs/2020A&A...644A..54S} {644, A54}

\bibitem[\protect\citeauthoryear{{Saxena}, {R{\"o}ttgering}  \&
  {Rigby}}{{Saxena} et~al.}{2017}]{saxena17}
{Saxena} A.,  {R{\"o}ttgering} H.~J.~A.,   {Rigby} E.~E.,  2017, \mn@doi
  [\mnras] {10.1093/mnras/stx1150}, \href
  {https://ui.adsabs.harvard.edu/abs/2017MNRAS.469.4083S} {469, 4083}

\bibitem[\protect\citeauthoryear{{Saxena} et~al.,}{{Saxena}
  et~al.}{2018}]{saxena18b}
{Saxena} A.,  et~al., 2018, \mn@doi [\mnras] {10.1093/mnras/sty152}, \href
  {https://ui.adsabs.harvard.edu/abs/2018MNRAS.475.5041S} {475, 5041}

\bibitem[\protect\citeauthoryear{{Saxena} et~al.,}{{Saxena}
  et~al.}{2019}]{saxena19}
{Saxena} A.,  et~al., 2019, \mn@doi [\mnras] {10.1093/mnras/stz2516}, \href
  {https://ui.adsabs.harvard.edu/abs/2019MNRAS.489.5053S} {489, 5053}

\bibitem[\protect\citeauthoryear{{Scharf}, {Smail}, {Ivison}, {Bower}, {van
  Breugel}  \& {Reuland}}{{Scharf} et~al.}{2003}]{scharf03}
{Scharf} C.,  {Smail} I.,  {Ivison} R.,  {Bower} R.,  {van Breugel} W.,
  {Reuland} M.,  2003, \mn@doi [\apj] {10.1086/377531}, \href
  {https://ui.adsabs.harvard.edu/abs/2003ApJ...596..105S} {596, 105}

\bibitem[\protect\citeauthoryear{{Schmidt} et~al.,}{{Schmidt}
  et~al.}{1998}]{schmidt98}
{Schmidt} M.,  et~al., 1998, \mn@doi [\aap] {10.48550/arXiv.astro-ph/9709144},
  \href {https://ui.adsabs.harvard.edu/abs/1998A&A...329..495S} {329, 495}

\bibitem[\protect\citeauthoryear{{Smail} \& {Blundell}}{{Smail} \&
  {Blundell}}{2013}]{smail13}
{Smail} I.,  {Blundell} K.~M.,  2013, \mn@doi [\mnras] {10.1093/mnras/stt1240},
  \href {https://ui.adsabs.harvard.edu/abs/2013MNRAS.434.3246S} {434, 3246}

\bibitem[\protect\citeauthoryear{{Swinbank} et~al.,}{{Swinbank}
  et~al.}{2015}]{swinbank15}
{Swinbank} A.~M.,  et~al., 2015, \mn@doi [\mnras] {10.1093/mnras/stv366}, \href
  {https://ui.adsabs.harvard.edu/abs/2015MNRAS.449.1298S} {449, 1298}

\bibitem[\protect\citeauthoryear{{Tozzi} \& {Norman}}{{Tozzi} \&
  {Norman}}{2001}]{tozzi01}
{Tozzi} P.,  {Norman} C.,  2001, \mn@doi [\apj] {10.1086/318237}, \href
  {https://ui.adsabs.harvard.edu/abs/2001ApJ...546...63T} {546, 63}

\bibitem[\protect\citeauthoryear{{Veilleux} \& {Osterbrock}}{{Veilleux} \&
  {Osterbrock}}{1987}]{veilleux87}
{Veilleux} S.,  {Osterbrock} D.~E.,  1987, \mn@doi [\apjs] {10.1086/191166},
  \href {https://ui.adsabs.harvard.edu/abs/1987ApJS...63..295V} {63, 295}

\bibitem[\protect\citeauthoryear{{Venemans} et~al.,}{{Venemans}
  et~al.}{2002}]{venemans02}
{Venemans} B.~P.,  et~al., 2002, \mn@doi [\apjl] {10.1086/340563}, \href
  {https://ui.adsabs.harvard.edu/abs/2002ApJ...569L..11V} {569, L11}

\bibitem[\protect\citeauthoryear{{Venemans} et~al.,}{{Venemans}
  et~al.}{2007}]{venemans07}
{Venemans} B.~P.,  et~al., 2007, \mn@doi [\aap] {10.1051/0004-6361:20053941},
  \href {https://ui.adsabs.harvard.edu/abs/2007A&A...461..823V} {461, 823}

\bibitem[\protect\citeauthoryear{{Villa-V{\'e}lez}, {Buat}, {Theul{\'e}},
  {Boquien}  \& {Burgarella}}{{Villa-V{\'e}lez} et~al.}{2021}]{villa-velez21}
{Villa-V{\'e}lez} J.~A.,  {Buat} V.,  {Theul{\'e}} P.,  {Boquien} M.,
  {Burgarella} D.,  2021, \mn@doi [\aap] {10.1051/0004-6361/202140890}, \href
  {https://ui.adsabs.harvard.edu/abs/2021A&A...654A.153V} {654, A153}

\bibitem[\protect\citeauthoryear{{Villar-Mart{\'\i}n}, {Tadhunter}, {Morganti},
  {Axon}  \& {Koekemoer}}{{Villar-Mart{\'\i}n} et~al.}{1999}]{villar1999}
{Villar-Mart{\'\i}n} M.,  {Tadhunter} C.,  {Morganti} R.,  {Axon} D.,
  {Koekemoer} A.,  1999, \mn@doi [\mnras] {10.1046/j.1365-8711.1999.02603.x},
  \href {https://ui.adsabs.harvard.edu/abs/1999MNRAS.307...24V} {307, 24}

\bibitem[\protect\citeauthoryear{{Villar-Mart{\'\i}n}, {Humphrey}, {De Breuck},
  {Fosbury}, {Binette}  \& {Vernet}}{{Villar-Mart{\'\i}n}
  et~al.}{2007a}]{villar-martin07b}
{Villar-Mart{\'\i}n} M.,  {Humphrey} A.,  {De Breuck} C.,  {Fosbury} R.,
  {Binette} L.,   {Vernet} J.,  2007a, \mn@doi [\mnras]
  {10.1111/j.1365-2966.2006.11371.x}, \href
  {https://ui.adsabs.harvard.edu/abs/2007MNRAS.375.1299V} {375, 1299}

\bibitem[\protect\citeauthoryear{{Villar-Mart{\'\i}n}, {S{\'a}nchez},
  {Humphrey}, {Dijkstra}, {di Serego Alighieri}, {De Breuck}  \& {Gonz{\'a}lez
  Delgado}}{{Villar-Mart{\'\i}n} et~al.}{2007b}]{villar-martin07a}
{Villar-Mart{\'\i}n} M.,  {S{\'a}nchez} S.~F.,  {Humphrey} A.,  {Dijkstra} M.,
  {di Serego Alighieri} S.,  {De Breuck} C.,   {Gonz{\'a}lez Delgado} R.,
  2007b, \mn@doi [\mnras] {10.1111/j.1365-2966.2007.11811.x}, \href
  {https://ui.adsabs.harvard.edu/abs/2007MNRAS.378..416V} {378, 416}

\bibitem[\protect\citeauthoryear{{Wang} et~al.,}{{Wang}
  et~al.}{2024}]{wang2023}
{Wang} W.,  et~al., 2024, \mn@doi [arXiv e-prints] {10.48550/arXiv.2401.02479},
  \href {https://ui.adsabs.harvard.edu/abs/2024arXiv240102479W} {p.
  arXiv:2401.02479}

\bibitem[\protect\citeauthoryear{{Welch}, {Rigby}  \& {Hutchison}}{{Welch}
  et~al.}{2023}]{welch23}
{Welch} B.,  {Rigby} J.~R.,   {Hutchison} T.~A.,  2023, \mn@doi [Research Notes
  of the American Astronomical Society] {10.3847/2515-5172/acb686}, \href
  {https://ui.adsabs.harvard.edu/abs/2023RNAAS...7...17W} {7, 17}

\bibitem[\protect\citeauthoryear{{Williams} et~al.,}{{Williams}
  et~al.}{2018}]{williams18}
{Williams} W.~L.,  et~al., 2018, \mn@doi [\mnras] {10.1093/mnras/sty026}, \href
  {https://ui.adsabs.harvard.edu/abs/2018MNRAS.475.3429W} {475, 3429}

\bibitem[\protect\citeauthoryear{{Xu}, {Livio}  \& {Baum}}{{Xu}
  et~al.}{1999}]{xu99}
{Xu} C.,  {Livio} M.,   {Baum} S.,  1999, \mn@doi [\aj] {10.1086/301007}, \href
  {https://ui.adsabs.harvard.edu/abs/1999AJ....118.1169X} {118, 1169}

\bibitem[\protect\citeauthoryear{{Zirm} et~al.,}{{Zirm} et~al.}{2005}]{zirm05}
{Zirm} A.~W.,  et~al., 2005, \mn@doi [\apj] {10.1086/431921}, \href
  {https://ui.adsabs.harvard.edu/abs/2005ApJ...630...68Z} {630, 68}

\bibitem[\protect\citeauthoryear{{van Dokkum}}{{van Dokkum}}{2001}]{lacosmic}
{van Dokkum} P.~G.,  2001, \mn@doi [\pasp] {10.1086/323894}, \href
  {https://ui.adsabs.harvard.edu/abs/2001PASP..113.1420V} {113, 1420}

\bibitem[\protect\citeauthoryear{{van Ojik}, {Roettgering}, {Carilli}, {Miley},
  {Bremer}  \& {Macchetto}}{{van Ojik} et~al.}{1996}]{vanojik96}
{van Ojik} R.,  {Roettgering} H.~J.~A.,  {Carilli} C.~L.,  {Miley} G.~K.,
  {Bremer} M.~N.,   {Macchetto} F.,  1996, \mn@doi [\aap]
  {10.48550/arXiv.astro-ph/9608099}, \href
  {https://ui.adsabs.harvard.edu/abs/1996A&A...313...25V} {313, 25}

\makeatother
\end{thebibliography}






\bsp	
\label{lastpage}
\end{document}